\documentclass[10pt,thmsa]{article}
\usepackage{amssymb}
\input{tcilatex}
\begin{document}
\begin{center}
{\Large Functional Integral Method in Quantum Theory }
\vskip 0.2cm
{\Large of Composite Particles and Quasiparticles}{\footnote{%
Lectures at the 7-th International School in Theoretical Physics, Hanoi 23
December 2000-05 January 2001.}}

\vskip 0.4cm
\textbf{Nguyen Van Hieu}

\textit{Institute of Physics \& Institute of Materials Science, NCST of
Vietnam }

and

\textit{Faculty of Technology, Vietnam National University, Hanoi}

\vskip 1cm
\textbf{Abstract}
\end{center}

\begin{quotation}
In this series of lectures we present the universal method based on the use
of the functional integrals to derive the bound state equations for
different two-body systems in elementary particle physics as well as in
condensed matter theory: the relativistic Nambu-Jona-Lasinio equation and
the relativistic Bethe-Salpeter equation in elementary particle physics, the
Bethe-Salpeter equation for the bipolaritons. The bound state equation for
the three-body systems is also established.
\end{quotation}

\section{Introduction}

The theoretical study of the bound states of different systems of
fundamental particles, their energy (mass) spectra and the corresponding
many-body quantum-mechanical wave functions or quantum field-theoretical
state vectors was always one among the main problems of the quantum physics.
The simplest examples of the two-body quantum systems are the hydrogen atom
and the positronium. In the non-relativistic approximation for the proton
the energy spectrum and the wave functions of the hydrogen atom are
determined by the Dirac equation with the Coulomb potential of the
attractive interaction between the electron and the proton. For the
formation of the bound states of two particles (including
particle-antiparticle pairs) due to the boson exchange in the relativistic
quantum field theory, for example the positronium in QED, we have the
Bethe-Salpeter equation$^{[1]}$. This relativistic integral equation was
widely applied to study the bound states of different two-body systems: the
mesons as the bound states of the quark-antiquark pairs in QCD$^{\left[
2-4\right] }$ and the diquarks as the bound states of the quark-quark pairs
in dense QCD$^{\left[ 5,6\right] }$. The direct four-fermion coupling may be
also the physical origin of the formation of the composite particles as the
bound states of some fermion-antifermion pairs or two-fermion systems. In
this case we have the Nambu-Jona-Lasinio equation$^{\left[ 7\right] }$. It
was widely applied to the study of the formation and the physical properties
of the mesons as the bound states of the quark-antiquark pairs$^{\left[
8-15\right] }$ . Recently this equation was used for the investigation of
the instanton induced quark-quark pairing in dense QCD$^{\left[ 16-21\right]
}$. In the study of the formation of the bound states of two-body systems of
quasiparticles in condensed matters one can use the Schrodinger equation if
the effective mass approximation is valid. However, there are quasiparticles
with the complicated energy spectra so that we cannot establish the
Schrodinger equation. In this case we must derive and then apply either the
Bethe-Salpeter or the Nambu-Jona-Lasinio equation. For example, due to the
exciton-exciton interaction there may exist the bound states of the systems
of two polaritons called the bipolaritons. The Bethe-Salpeter equation for
the bipolaritons was established and studied by many authors$^{\left[
22,23\right] }$.

The relativistic Bethe-Salpeter equation for the systems of two elementary
particles as well as the nonrelativistic Bethe-Salpeter equation for the
systems of two quasiparticles with the complicated structure of the energy
spectra in condensed matters were derived within the framework of the
perturbation theory by summing up the matrix elements of the infinite series
of Feynman diagrams in the ladder approximation. The Nambu-Jona-Lasinio
equation was established in the pioneering work of Nambu and Jona-Lasinio
also by means of the perturbation theory$^{\left[ 7\right] }$. Subsequently
the elegant derivation of this famous equation by means of the functional
integral technique was given by Eguchi, Sugawara$^{\left[ 8,9\right] }$ and
Kikkawa$^{\left[ 10\right] }$. Within the functional integral formalism it
is not necessary to sum up the infinite series of the matrix elements of the
Feynman diagrams in the ladder approximation, and the derivation of the
equations is simplified. In the attempt to present the functional integral
technique as the universal method for the study of many different problems
in the quantum theory$^{\left[ 24-29\right] }$ in this series of lectures we
review the derivation of both types of bound state equations in the
elementary particle physics as well as in the condensed matter theory by
means of the functional integral technique. The derivation of the bound
state equations for the systems of two quasiparticles in condensed matters
by means of the functional integral method is presented first time in these
lectures. As we shall see, it is straightforward to establish the bound
state equations for the three-body systems$^{\left[ 30\right] }$ in the
framework of the functional integral formalism.

\section{Relativistic Nambu-Jona-Lasinio Equation}

We start to study the equations for the bound states of many-body systems by
considering the simplest examples : the formation of the mesons and the
diquarks due to the direct $4$-fermion coupling of the quark field with the
interaction Lagrangian

\begin{eqnarray}
L_{\text{int}}\left( x\right) &=&\frac{1}{2}\overline{\psi }^{A}\left(
x\right) \psi _{B}\left( x\right) U_{AC}^{BD}\overline{\psi }^{C}\left(
x\right) \psi _{D}\left( x\right) ,  \label{1} \\
U_{AC}^{BD} &=&-U_{CA}^{BD}=-U_{AC}^{DB}=U_{CA}^{DB},  \nonumber
\end{eqnarray}

\noindent where $\psi _{A}\left( x\right) $ \noindent denote the quark field,

\[
A=\left( \alpha ai\right) ,\qquad B=\left( \beta bj\right) ,\qquad C=\left(
\gamma ck\right) ,\qquad D=\left( \delta dl\right) 
\]

\noindent are the sets consisting of the Dirac spinor indices $\alpha ,\beta
,\gamma ,\delta =1,2,3,4,$ \noindent the indices of the color symmetry $%
a,b,c,d=1,2,...N_{c}$ \noindent and those of the flavor one $%
i,j,k,l=1,2,...N_{f}$. \noindent For the convenience in the study of the
diquark formation we write the interaction Lagrangian (1) also in another
form

\begin{equation}
L_{\text{int}}\left( x\right) =\frac{1}{2}\overline{\psi }^{A}\left(
x\right) \overline{\psi }^{C}\left( x\right) V_{CA}^{BD}\psi _{D}\left(
x\right) \psi _{B}\left( x\right) ,  \label{2}
\end{equation}

\noindent with the new coupling constants 
\begin{equation}
V_{CA}^{BD}=U_{AC}^{BD}.  \label{3}
\end{equation}

\noindent The mathematical tool in the functional integral formalism is the
functional integral of the field system

\begin{equation}
Z=\int \left[ D\psi \right] \left[ D\overline{\psi }\right] \exp \left\{
iS\left[ \psi ,\overline{\psi }\right] \right\} ,  \label{4}
\end{equation}

\noindent where $S\left[ \psi ,\overline{\psi }\right] $ \noindent is the
total action of the system

\begin{equation}
S\left[ \psi ,\overline{\psi }\right] =S_{0}\left[ \psi ,\overline{\psi }%
\right] +S_{\text{int}}\left[ \psi ,\overline{\psi }\right] ,  \label{5}
\end{equation}

\noindent $S_{0}\left[ \psi ,\overline{\psi }\right] $ being the action of
the free quark field with some bare mass $m$

\begin{eqnarray}
S_{0}\left[ \psi ,\overline{\psi }\right] &=&-\int d^{4}x\overline{\psi }%
^{A}\left( x\right) \left[ \left( \widehat{\partial }\right)
_{A}^{B}+m\delta _{A}^{B}\right] \psi _{B}\left( x\right) ,  \label{6} \\
\left( \widehat{\partial }\right) _{A}^{B} &=&\delta _{a}^{b}\delta
_{i}^{j}\left( \gamma _{\mu }\right) _{\,\alpha }^{\beta }\partial ^{\mu } 
\nonumber
\end{eqnarray}

\noindent and $S_{\text{int}}\left[ \psi ,\overline{\psi }\right] $
\noindent being the contribution of the interaction Lagrangian to the total
action

\begin{equation}
S_{\text{int}}\left[ \psi ,\overline{\psi }\right] =\int d^{4}xL_{\text{int}%
}\left( x\right) .  \label{7}
\end{equation}
\smallskip \noindent For the study of the mesons as the bound states of the
quark-antiquark pairs we use the interaction Lagrangian in the form (1),
introduce the composite meson field $\Phi _{B}^{A}\left( x\right) $
\noindent and set

\begin{equation}
Z_{0}^{\Phi }=\int \left[ D\Phi \right] \exp \left\{ -\frac{i}{2}\int
d^{4}x\Phi _{B}^{A}\left( x\right) U_{AC}^{BD}\Phi _{D}^{C}\left( x\right)
\right\} .  \label{8}
\end{equation}

\noindent Shifting the functional integration variables

\[
\Phi _{B}^{A}\left( x\right) \rightarrow \Phi _{B}^{A}\left( x\right) +%
\overline{\psi }^{A}\left( x\right) \psi _{B}\left( x\right) , 
\]

\noindent we establish the Hubbard-Stratonovich transformation

\begin{eqnarray}
&&\exp \left\{ \frac{i}{2}\int d^{4}x\overline{\psi }^{A}\left( x\right)
\psi _{B}\left( x\right) U_{AC}^{BD}\overline{\psi }^{C}\left( x\right) \psi
_{D}\left( x\right) \right\} 
\begin{array}{l}
=
\end{array}
\label{9} \\
&=&\frac{1}{Z_{0}^{\Phi }}\int \left[ D\Phi \right] \exp \left\{ -\frac{i}{2}%
\int d^{4}x\Phi _{B}^{A}\left( x\right) U_{AC}^{BD}\Phi _{D}^{C}\left(
x\right) \right\} \exp \left\{ -i\int d^{4}x\overline{\psi }^{A}\left(
x\right) \psi _{B}\left( x\right) \Delta _{A}^{B}\left( x\right) \right\} 
\nonumber
\end{eqnarray}

\noindent with

\begin{equation}
\Delta _{A}^{B}\left( x\right) =U_{AC}^{BD}\Phi _{D}^{C}\left( x\right)
\label{10}
\end{equation}

\noindent Using this formula, we transform the expression (4) of the
functional integral of the quark field into that of the composite meson one

\begin{equation}
Z=\frac{Z_{0}}{Z_{0}^{\Phi }}\int \left[ D\Phi \right] \exp \left\{ iS_{%
\text{eff}}\left[ \Phi \right] \right\}  \label{11}
\end{equation}

\noindent with the effective action

\begin{equation}
S_{\text{eff}}\left[ \Phi \right] =-\frac{1}{2}\int d^{4}x\Phi
_{B}^{A}\left( x\right) U_{AC}^{BD}\Phi _{D}^{C}\left( x\right) +W\left[
\Delta \right] ,  \label{12}
\end{equation}

\noindent where

\begin{equation}
Z_{0}=\int \left[ D\psi \right] \left[ D\overline{\psi }\right] \exp \left\{
iS_{0}\left[ \psi ,\overline{\psi }\right] \right\}  \label{13}
\end{equation}

\noindent and the functional $W\left[ \Delta \right] $ is determined by the
formula

\begin{eqnarray}
\exp \left\{ iW\left[ \Delta \right] \right\} &=&\frac{1}{Z_{0}}\int \left[
D\psi \right] \left[ D\overline{\psi }\right] \exp \left\{ iS_{0}\left[ \psi
,\overline{\psi }\right] \right\}  \nonumber \\
&&\exp \left\{ -i\int d^{4}x\overline{\psi }^{A}\left( x\right) \psi
_{B}\left( x\right) \Delta _{A}^{B}\left( x\right) \right\} .  \label{14}
\end{eqnarray}

\noindent We can express $W\left[ \Delta \right] $ in the form of a
functional power series in the field $\Delta _{A}^{B}\left( x\right) $

\begin{equation}
W\left[ \Delta \right] =\sum\limits_{n=1}^{\infty }W^{\left( n\right)
}\left[ \Delta \right] ,  \label{15}
\end{equation}

\noindent where $W^{\left( n\right) }\left[ \Delta \right] $ is a
homogeneous functional of the $n$-th order. Calculations give, for examples,

\begin{equation}
W^{\left( 1\right) }\left[ \Delta \right] =-i\int d^{4}xS_{A}^{B}\left(
0\right) \Delta _{B}^{A}\left( x\right) ,  \label{16}
\end{equation}

\begin{equation}
W^{\left( 2\right) }\left[ \Delta \right] =\frac{i}{2}\int d^{4}x\int
d^{4}yS_{A}^{B}\left( x-y\right) \Delta _{B}^{C}\left( y\right)
S_{C}^{D}\left( y-x\right) \Delta _{D}^{A}\left( x\right) ,  \label{17}
\end{equation}

\begin{equation}
W^{\left( 3\right) }\left[ \Delta \right] =-\frac{i}{3}\int d^{4}x\int
d^{4}y\int d^{4}zS_{A}^{B}\left( x-y\right) \Delta _{B}^{C}\left( y\right)
S_{C}^{D}\left( y-z\right) \Delta _{D}^{E}\left( z\right) S_{E}^{A}\left(
z-x\right) ,  \label{18}
\end{equation}

\noindent etc, where $S_{A}^{B}\left( x-y\right) $ is the two-point Green
function of the free quark field

\begin{equation}
S_{A}^{B}\left( x-y\right) =\frac{i}{Z_{0}}\int \left[ D\psi \right] \left[ D%
\overline{\psi }\right] \psi _{A}\left( x\right) \overline{\psi }^{B}\left(
y\right) \exp \left\{ iS_{0}\left[ \psi ,\overline{\psi }\right] \right\} .
\label{19}
\end{equation}

\noindent It satisfies the equation

\begin{equation}
\left[ \left( \widehat{\partial }\right) _{A}^{B}+m\delta _{A}^{B}\right]
S_{B}^{C}\left( x-y\right) =\delta _{A}^{C}\delta \left( x-y\right) .
\label{20}
\end{equation}

\noindent Denote $\widetilde{S}_{A}^{B}\left( \mathbf{p}\right) $ its
Fourier transform

\begin{equation}
S_{A}^{B}\left( x\right) =\frac{1}{\left( 2\pi \right) ^{4}}\int
d^{4}p\,e^{ipx}\widetilde{S}_{A}^{B}\left( \mathbf{p}\right) .  \label{21}
\end{equation}

\noindent We have

\begin{equation}
\widetilde{S}_{A}^{B}\left( \mathbf{p}\right) =\delta _{a}^{b}\delta
_{i}^{j}\left( \frac{1}{i\widehat{p}+m}\right) _{\alpha }^{\beta }.
\label{22}
\end{equation}

From the variational principle

\begin{equation}
\frac{\delta S_{\text{eff}}\left[ \Phi \right] }{\delta \Phi _{D}^{C}\left(
x\right) }=0  \label{23}
\end{equation}

\noindent we derive the field equation

\begin{equation}
\Delta _{C}^{D}\left( x\right) =U_{CA}^{DB}\frac{\delta W\left[ \Delta
\right] }{\delta \Delta _{A}^{B}\left( x\right) }.  \label{24}
\end{equation}

\noindent For some definite types of composite meson fields the first order
functional $W^{\left( 1\right) }\left[ \Delta \right] \,$either vanishes or
does not give the contribution. Then in the second order with respect to $%
\Delta _{A}^{B}\left( x\right) $ $\,$we can replace $W\left[ \Delta \right] $
by $W^{\left( 2\right) }\left[ \Delta \right] $ and obtain the approximate
field equation

\begin{equation}
\Delta _{C}^{D}\left( x\right) =U_{CA}^{DB}\frac{\delta W^{\left( 2\right)
}\left[ \Delta \right] }{\delta \Delta _{A}^{B}\left( x\right) }.  \label{25}
\end{equation}

\noindent Substituting the expression (17) of $W^{\left( 2\right) }\left[
\Delta \right] $ \noindent into the r.h.s. of the relation (25), we obtain
immediately the Nambu-Jona-Lasinio equation

\begin{equation}
\Delta _{C}^{D}\left( x\right) =iU_{CA}^{DB}\int d^{4}yS_{B}^{E}\left(
x-y\right) \Delta _{E}^{F}\left( y\right) S_{F}^{A}\left( y-x\right) .
\label{26}
\end{equation}

\noindent For establishing the field equation in the general case with
non-vanishing $W^{\left( 1\right) }\left[ \Delta \right] $ \noindent we must
take into account the contributions of the functionals $W^{\left( n\right)
}\left[ \Delta \right] $ of all orders by summing up the infinite series
(15). Denote

\begin{eqnarray}
\mathbf{S}_{A}^{B}\left( x,y\right) &=&S_{A}^{B}\left( x-y\right) -\int
d^{4}x_{1}S_{A}^{B_{1}}\left( x-x_{1}\right) \Delta _{B_{1}}^{A_{1}}\left(
x_{1}\right) S_{A_{1}}^{B}\left( x_{1}-y\right)  \nonumber \\
&&+\int d^{4}x_{1}\int d^{4}x_{2}S_{A}^{B_{1}}\left( x-x_{1}\right) \Delta
_{B_{1}}^{A_{1}}\left( x_{1}\right) S_{A_{1}}^{B_{2}}\left(
x_{1}-x_{2}\right) \Delta _{B_{2}}^{A_{2}}\left( x_{2}\right)
S_{A_{2}}^{B}\left( x_{2}-y\right)  \nonumber \\
&&-\int d^{4}x_{1}\int d^{4}x_{2}\int d^{4}x_{3}S_{A}^{B_{1}}\left(
x-x_{1}\right) \Delta _{B_{1}}^{A_{1}}\left( x_{1}\right)
S_{A_{1}}^{B_{2}}\left( x_{1}-x_{2}\right)  \nonumber \\
&&\Delta _{B_{2}}^{A_{2}}\left( x_{2}\right) S_{A_{2}}^{B_{3}}\left(
x_{2}-x_{3}\right) \Delta _{B_{3}}^{A_{3}}\left( x_{3}\right)
S_{A_{3}}^{B}\left( x_{3}-y\right) +.....  \label{27}
\end{eqnarray}

\noindent the two-point Green function of the quark field interacting with
the composite meson field $\Delta _{B}^{A}\left( x\right) .$ It is
determined by the Schwinger-Dyson equation

\begin{equation}
\mathbf{S}_{A}^{B}\left( x,y\right) =S_{A}^{B}\left( x-y\right) -\int
d^{4}zS_{A}^{C}\left( x-z\right) \Delta _{C}^{D}\left( z\right) \mathbf{S}%
_{D}^{B}\left( z,y\right) .  \label{28}
\end{equation}

\noindent In terms of this new two-point Green function we can rewrite the
field equation $(24)$ in the form

\begin{equation}
\Delta _{C}^{D}\left( x\right) +iU_{CA}^{DB}\mathbf{S}_{B}^{A}\left(
x,x\right) =0.  \label{29}
\end{equation}

\noindent In general there exists some non-vanishing constant solution of
the nonlinear equation (29)

\begin{eqnarray}
\left( \Phi ^{0}\right) _{A}^{B} &=&\text{const,}  \nonumber \\
\left( \Delta ^{0}\right) _{A}^{B} &=&U_{AC}^{BD}\left( \Phi ^{0}\right)
_{D}^{C}=\text{const.}  \label{30}
\end{eqnarray}

\noindent With the constant meson field the expression in the r.h.s. of the
formula (27) depends only on the difference $x-y$ of the coordinates $x$ and 
$y$

\begin{equation}
\mathbf{S}_{B}^{A}\left( x,y\right) =\mathbf{S}_{B}^{A}\left( x-y\right) ,
\label{31}
\end{equation}

\noindent and this constant meson field is determined by the equation

\begin{equation}
\left( \Delta ^{0}\right) _{C}^{D}+iU_{CA}^{DB}\mathbf{S}_{B}^{A}\left(
0\right) =0.  \label{32}
\end{equation}

\noindent Consider the quantum fluctuations of the meson field around the
background constant one (30) satisfying the equation (32) and set

\begin{eqnarray}
\Phi _{A}^{B}\left( x\right) &=&\left( \Phi ^{0}\right) _{B}^{A}+\varphi
_{B}^{A}\left( x\right) ,  \nonumber \\
\Delta _{B}^{A}\left( x\right) &=&\left( \Delta ^{0}\right) _{B}^{A}+\xi
_{B}^{A}\left( x\right) ,  \label{33}
\end{eqnarray}

\noindent with 
\begin{equation}
\xi _{B}^{A}\left( x\right) =U_{BD}^{AC}\varphi _{C}^{D}\left( x\right) .
\label{34}
\end{equation}

\noindent Up to the second order with respect to the quantum fluctuations
the effective action of the composite meson field equals

\begin{equation}
S_{\text{eff}}\left[ \Phi \right] \approx S_{\text{eff}}\left[ \Phi
^{0}\right] +\frac{1}{2}\int d^{4}x\int d^{4}y\varphi _{B}^{A}\left(
x\right) \frac{\delta ^{2}S_{\text{eff}}\left[ \Phi \right] }{\delta \Phi
_{B}^{A}\left( x\right) \delta \Phi _{D}^{C}\left( y\right) }|_{_{\Phi =\Phi
^{0}}}\varphi _{D}^{C}\left( y\right) .  \label{35}
\end{equation}

\noindent Using the expression (12) of $S_{\text{eff}}\left[ \Phi \right]
,\, $we have

\begin{eqnarray}
S_{\text{eff}}\left[ \Phi \right] &\approx &S_{\text{eff}}\left[ \Phi
^{0}\right] -\frac{1}{2}\int d^{4}x\varphi _{B}^{A}\left( x\right)
U_{AC}^{BD}\varphi _{D}^{C}\left( x\right)  \nonumber \\
&&+\frac{1}{2}\int d^{4}x\int d^{4}y\xi _{B}^{A}\left( x\right) \frac{\delta
^{2}W\left[ \Delta \right] }{\delta \Delta _{B}^{A}\left( x\right) \delta
\Delta _{D}^{C}\left( y\right) }|_{_{\Delta =\Delta ^{0}}}\xi _{D}^{C}\left(
y\right) .  \label{36}
\end{eqnarray}

\noindent By summing up the infinite series

\[
\sum\limits_{n=1}^{\infty }\frac{\delta ^{2}W^{\left( n\right) }\left[
\Delta \right] }{\delta \Delta _{B}^{A}\left( x\right) \delta \Delta
_{D}^{C}\left( y\right) }|_{_{\Delta =\Delta ^{0}}} 
\]

\noindent we obtain

\begin{equation}
\frac{\delta ^{2}W\left[ \Delta \right] }{\delta \Delta _{B}^{A}\left(
x\right) \delta \Delta _{D}^{C}\left( y\right) }|_{_{\Delta =\Delta ^{0}}}=i%
\mathbf{S}_{B}^{C}\left( x-y\right) \mathbf{S}_{D}^{A}\left( y-x\right) .
\label{37}
\end{equation}

\noindent Then from the formula (36) for the effective action and the
variational principle it follows the Nambu-Jona-Lasinio equation containing
the two-point Green function of the quarks interacting with the constant
background meson field

\begin{equation}
\xi _{B}^{A}\left( x\right) =iU_{AC}^{BD}\int d^{4}y\mathbf{S}_{D}^{E}\left(
x-y\right) \xi _{E}^{F}\left( y\right) \mathbf{S}_{F}^{C}\left( y-x\right) .
\label{38}
\end{equation}

\noindent In the special case when the background meson field vanishes or
its contribution is neglected then the equation (38) reduces to the equation
(26).

Now we study the formation of the bound states of a pair of two quarks. For
this purpose we use the interaction Lagrangian in the form (2), introduce
the composite bosonic diquark field $\Phi _{AB}\left( x\right) \,$as well as
its conjugate $\overline{\Phi }^{BA}\left( x\right) $ and set

\begin{equation}
Z_{0}^{\Phi ,\overline{\Phi }}=\int \left[ D\Phi \right] \left[ D\overline{%
\Phi }\right] \exp \left\{ -\frac{i}{2}\int d^{4}x\overline{\Phi }%
^{AC}\left( x\right) V_{CA}^{BD}\Phi _{DB}\left( x\right) \right\} .
\label{39}
\end{equation}

\noindent Shifting the functional integration variables

\begin{eqnarray}
\Phi _{DB}\left( x\right) &\rightarrow &\Phi _{DB}\left( x\right) +\psi
_{D}\left( x\right) \psi _{B}\left( x\right) ,  \nonumber \\
\overline{\Phi }^{AC}\left( x\right) &\rightarrow &\overline{\Phi }%
^{AC}\left( x\right) +\overline{\psi }^{A}\left( x\right) \overline{\psi }%
^{C}\left( x\right) ,  \label{40}
\end{eqnarray}

\noindent we establish the Hubbard-Stratonovich transformation

\begin{eqnarray}
&&\exp \left\{ \frac{i}{2}\int d^{4}x\overline{\psi }^{A}\left( x\right) 
\overline{\psi }^{C}\left( x\right) V_{CA}^{BD}\psi _{D}\left( x\right) \psi
_{B}\left( x\right) \right\} 
\begin{array}{l}
=
\end{array}
\nonumber \\
&=&\frac{1}{Z_{0}^{\Phi ,\overline{\Phi }}}\int \left[ D\Phi \right] \left[ D%
\overline{\Phi }\right] \exp \left\{ -\frac{i}{2}\int d^{4}x\overline{\Phi }%
^{AC}\left( x\right) V_{CA}^{BD}\Phi _{DB}\left( x\right) \right\}
\label{41} \\
&&.\exp \left\{ -\frac{i}{2}\int d^{4}x\overline{\psi }^{A}\left( x\right) 
\overline{\psi }^{C}\left( x\right) \Delta _{CA}\left( x\right) +\overline{%
\Delta }^{BD}\left( x\right) \psi _{D}\left( x\right) \psi _{B}\left(
x\right) \right\} ,  \nonumber
\end{eqnarray}

\noindent where

\begin{eqnarray}
\Delta _{CA}\left( x\right) &=&V_{CA}^{BD}\Phi _{DB}\left( x\right) , 
\nonumber \\
\overline{\Delta }^{BD}\left( x\right) &=&\overline{\Phi }^{AC}\left(
x\right) V_{CA}^{BD}.  \label{42}
\end{eqnarray}

\noindent Using formula (41), we can transform the expression (4) of the
functional integral of the quark field into that of the diquark system

\begin{equation}
Z=\frac{Z^{0}}{Z_{0}^{\Phi ,\overline{\Phi }}}\int \left[ D\Phi \right]
\left[ D\overline{\Phi }\right] \exp \left\{ iS_{\text{eff}}\left[ \Phi ,%
\overline{\Phi }\right] \right\}  \label{43}
\end{equation}

\noindent with the effective action

\begin{equation}
S_{\text{eff}}\left[ \Phi ,\overline{\Phi }\right] =-\frac{1}{2}\int d^{4}x%
\overline{\Phi }^{AC}\left( x\right) V_{CA}^{BD}\Phi _{DB}\left( x\right)
+W\left[ \Delta ,\overline{\Delta }\right] ,  \label{44}
\end{equation}

\noindent $W\left[ \Delta ,\overline{\Delta }\right] $ being determined by
the formula

\begin{eqnarray}
&&\exp \left\{ iW\left[ \Delta ,\overline{\Delta }\right] \right\} 
\begin{array}{l}
=
\end{array}
\frac{1}{Z^{0}}\int \left[ D\psi \right] \left[ D\overline{\psi }\right]
\exp \left\{ iS_{0}\left[ \psi ,\overline{\psi }\right] \right\}  \label{45}
\\
&&\qquad \exp \left\{ -\frac{i}{2}\int d^{4}x\left[ \overline{\psi }%
^{A}\left( x\right) \overline{\psi }^{C}\left( x\right) \Delta _{CA}\left(
x\right) +\overline{\Delta }^{BD}\left( x\right) \psi _{D}\left( x\right)
\psi _{B}\left( x\right) \right] \right\} .\qquad  \nonumber
\end{eqnarray}

\noindent and having the form

\begin{equation}
W\left[ \Delta ,\overline{\Delta }\right] =\sum\limits_{n=1}^{\infty
}W^{\left( 2n\right) }\left[ \Delta ,\overline{\Delta }\right] ,  \label{46}
\end{equation}

\noindent where $W^{\left( 2n\right) }\left[ \Delta ,\overline{\Delta }%
\right] $ is a homogeneous functional of the $n$-th order with respect to
each type of bosonic fields $\Delta _{CA}\left( x\right) \,$and $\overline{%
\Delta }^{BD}\left( x\right) .$ \noindent Calculations give

\begin{equation}
W^{\left( 2\right) }\left[ \Delta ,\overline{\Delta }\right] =-\frac{i}{2}%
\int d^{4}x_{1}\int d^{4}x_{2}\overline{\Delta }^{A_{1}B_{1}}\left(
x_{1}\right) S_{A_{1}}^{A_{2}}\left( x_{1}-x_{2}\right)
S_{B_{1}}^{B_{2}}\left( x_{1}-x_{2}\right) \Delta _{B_{2}A_{2}}\left(
x_{2}\right) \qquad  \label{47}
\end{equation}

\noindent or in another form

\begin{equation}
W^{\left( 2\right) }\left[ \Delta ,\overline{\Delta }\right] =-\frac{i}{2}%
\int d^{4}x_{1}\int d^{4}x_{2}\overline{\Delta }^{A_{1}B_{1}}\left(
x_{1}\right) S_{B_{1}}^{B_{2}}\left( x_{1}-x_{2}\right) \Delta
_{B_{2}A_{2}}\left( x_{2}\right) S_{A_{1}}^{\text{T}A_{2}}\left(
x_{2}-x_{1}\right) \qquad  \label{48}
\end{equation}

\noindent convenient for the comparison with the higher order functionals

\begin{eqnarray}
W^{\left( 4\right) }\left[ \Delta ,\overline{\Delta }\right] &=&\frac{i}{4}%
\int d^{4}x_{1}...\int d^{4}x_{4}\overline{\Delta }^{A_{1}B_{1}}\left(
x_{1}\right) S_{B_{1}}^{B_{2}}\left( x_{1}-x_{2}\right) \Delta
_{B_{2}A_{2}}\left( x_{2}\right) S_{A_{3}}^{\text{T}A_{2}}\left(
x_{2}-x_{3}\right)  \nonumber \\
&&\qquad \qquad \overline{\Delta }^{A_{3}B_{3}}\left( x_{3}\right)
S_{B_{3}}^{B_{4}}\left( x_{3}-x_{4}\right) \Delta _{B_{4}A_{4}}\left(
x_{4}\right) S_{A_{1}}^{\text{T}A_{4}}\left( x_{4}-x_{1}\right) ,  \label{49}
\end{eqnarray}
\begin{eqnarray}
W^{\left( 6\right) }\left[ \Delta ,\overline{\Delta }\right] &=&-\frac{i}{6}%
\int d^{4}x_{1}...\int d^{4}x_{6}\overline{\Delta }^{A_{1}B_{1}}\left(
x_{1}\right) S_{B_{1}}^{B_{2}}\left( x_{1}-x_{2}\right) \Delta
_{B_{2}A_{2}}\left( x_{2}\right) S_{A_{3}}^{\text{T}A_{2}}\left(
x_{2}-x_{3}\right)  \nonumber \\
&&.......\,\,\,\,\,\overline{\Delta }^{A_{5}B_{5}}\left( x_{5}\right)
S_{B_{5}}^{B_{6}}\left( x_{5}-x_{6}\right) \Delta _{B_{6}A_{6}}\left(
x_{6}\right) S_{A_{1}}^{\text{T}A_{6}}\left( x_{6}-x_{1}\right) .  \label{50}
\end{eqnarray}

\noindent etc, where we used the notation

\begin{equation}
S_{A}^{\text{T}B}\left( y-x\right) =S_{A}^{B}\left( x-y\right) .  \label{51}
\end{equation}

\smallskip \noindent From the variational principle

\begin{equation}
\frac{\delta S_{\text{eff}}\left[ \Phi ,\overline{\Phi }\right] }{\delta 
\overline{\Phi }^{AB}\left( x\right) }=0  \label{52}
\end{equation}

\noindent we derive the field equation

\begin{equation}
\frac{1}{2}\Delta _{BA}\left( x\right) =V_{BA}^{CD}\frac{\delta W\left[
\Delta ,\overline{\Delta }\right] }{\delta \overline{\Delta }^{CD}\left(
x\right) }.  \label{53}
\end{equation}

\noindent If we neglect the contributions of the high order functionals $%
W^{\left( 2n\right) }\left[ \Delta ,\overline{\Delta }\right] $ with $n>1$
and use the approximation

\[
W\left[ \Delta ,\overline{\Delta }\right] \approx W^{\left( 2\right) }\left[
\Delta ,\overline{\Delta }\right] , 
\]

\noindent then the field equation (53) is the Nambu-Jona-Lasinio equation
for the composite diquark field

\begin{equation}
\Delta _{AB}\left( x\right) =-iV_{AB}^{DC}\int d^{4}yS_{C}^{E}\left(
x-y\right) S_{D}^{F}\left( x-y\right) \Delta _{EF}\left( y\right) .
\label{54}
\end{equation}
In order to take into account the contributions of the high order
functionals $W^{\left( 2n\right) }\left[ \Delta ,\overline{\Delta }\right] $
with $n>1$ we search the non-vanishing constant solution of the field
equation (53)

\begin{eqnarray}
\left( \Phi ^{0}\right) _{DB} &=&\text{const,\qquad }\left( \overline{\Phi }%
^{0}\right) ^{AC}=\text{const,}  \label{55} \\
\left( \Delta ^{0}\right) _{CA} &=&V_{CA}^{BD}\left( \Phi ^{0}\right) _{DB}=%
\text{const,\qquad }\left( \overline{\Delta }^{0}\right) ^{BD}=\left( 
\overline{\Phi }^{0}\right) ^{AC}V_{CA}^{BD}=\text{const}  \nonumber
\end{eqnarray}
and consider the quantum fluctuations around this background constant field :

\begin{eqnarray}
\Phi _{DB}\left( x\right) &=&\left( \Phi ^{0}\right) _{DB}+\varphi
_{DB}\left( x\right) ,\qquad \overline{\Phi }^{AC}\left( x\right) =\left( 
\overline{\Phi }^{0}\right) ^{AC}+\overline{\varphi }^{AC}\left( x\right) , 
\nonumber \\
\Delta _{CA}\left( x\right) &=&\left( \Delta ^{0}\right) _{CA}+\xi
_{CA}\left( x\right) ,\qquad \overline{\Delta }^{BD}\left( x\right) =\left( 
\overline{\Delta }^{0}\right) ^{BD}+\overline{\xi }^{BD}\left( x\right) ,
\label{56}
\end{eqnarray}
with

\begin{equation}
\xi _{CA}\left( x\right) =V_{CA}^{BD}\varphi _{DB}\left( x\right) ,\qquad 
\overline{\xi }^{BD}\left( x\right) =\overline{\varphi }^{AC}\left( x\right)
V_{CA}^{BD}.  \label{57}
\end{equation}
Up to the first order with respect to each type of bosonic fluctuations $%
\varphi _{DB}\left( x\right) ,\xi _{DB}\left( x\right) $ and $\overline{%
\varphi }^{AC}\left( x\right) ,\overline{\xi }^{AC}\left( x\right) $ we have
the effective action

\begin{eqnarray}
S_{\text{eff}}\left[ \Phi ,\overline{\Phi }\right] &\approx &S_{\text{eff}%
}\left[ \Phi ^{0},\overline{\Phi }^{0}\right] =\frac{1}{2}\int d^{4}x%
\overline{\varphi }^{AC}\left( x\right) V_{CA}^{BD}\varphi _{DB}\left(
x\right)  \label{58} \\
&&+\int d^{4}x\int d^{4}y\overline{\xi }^{AC}\left( x\right) \frac{\delta
^{2}W\left[ \Delta ,\overline{\Delta }\right] }{\delta \overline{\Delta }%
^{AC}\left( x\right) \delta \Delta _{DB}\left( y\right) }\vert_{\Delta =\Delta
^{0},\overline{\Delta }=\overline{\Delta }^{0}} \xi _{DB}\left(
x\right) .  \nonumber
\end{eqnarray}
By summing up the infinite series

\[
\sum\limits_{n=1}^{\infty }\frac{\delta ^{2}W\left[ \Delta ,\overline{\Delta 
}\right] }{\delta \overline{\Delta }^{AC}\left( x\right) \delta \Delta
_{DB}\left( y\right) } 
\]
we obtain

\begin{equation}
\frac{\delta ^{2}W\left[ \Delta ,\overline{\Delta }\right] }{\delta 
\overline{\Delta }^{AC}\left( x\right) \delta \Delta _{DB}\left( y\right) }
\vert_{\Delta =\Delta ^{0}, \overline{\Delta }=\overline{\Delta }^{0}} 
 =-\frac{i}{2}\mathbf{S}_{A}^{B}\left( x-y\right) \mathbf{S}%
_{C}^{D}\left( x-y\right) ,  \label{59}
\end{equation}
where $\mathbf{S}_{A}^{B}\left( x-y\right) $ is determined by the
Schwinger-Dyson equation

\begin{equation}
\mathbf{S}_{A}^{B}\left( x-y\right) =S_{A}^{B}\left( x-y\right) -\int
d^{4}z\int d^{4}uS_{A}^{C}\left( x-z\right) \left( \Delta ^{0}\right)
_{CD}S_{E}^{\text{T}D}\left( z-u\right) \left( \overline{\Delta }^{0}\right)
^{EF}\mathbf{S}_{F}^{B}\left( u-y\right) .\qquad  \label{60}
\end{equation}
Then from the expression (58) of the effective action and the variational
principle it follows the Nambu-Jona-Lasinio equation

\begin{equation}
\xi _{AB}\left( x\right) =-iV_{AB}^{DC}\int d^{4}y\mathbf{S}_{C}^{E}\left(
x-y\right) \mathbf{S}_{D}^{F}\left( x-y\right) \xi _{EF}\left( y\right)
\label{61}
\end{equation}
containing the two-point Green function $\mathbf{S}_{A}^{B}\left( x-y\right) 
$ of the quark field interacting with the background bosonic constant field
(55). In the special case of the vanishing background bosonic field $\left(
\Phi ^{0}\right) _{DB}=\left( \overline{\Phi }^{0}\right) ^{AC}=\left(
\Delta ^{0}\right) _{DB}=\left( \overline{\Delta }^{0}\right) ^{AC}=0$ the
equation (61) coincides with the equation (54).

In order to solve equation (60) we work in the momentum space. Denote $%
\widetilde{S}_{A}^{B}\left( p\right) $ and $\widetilde{\mathbf{S}}%
_{A}^{B}\left( p\right) $ the Fourier transforms of the two-point Green
functions $S_{A}^{B}\left( x\right) $ and $\mathbf{S}_{A}^{B}\left( x\right) 
$,

\begin{eqnarray}
S_{A}^{B}\left( x\right) &=&\frac{1}{\left( 2\pi \right) ^{4}}\int
d^{4}pe^{ipx}\widetilde{S}_{A}^{B}\left( p\right) ,  \nonumber \\
\mathbf{S}_{A}^{B}\left( x\right) &=&\frac{1}{\left( 2\pi \right) ^{4}}\int
d^{4}pe^{ipx}\widetilde{\mathbf{S}}_{A}^{B}\left( p\right) ,  \label{62} \\
px &=&\mathbf{px}-p_{0}x_{0}=\mathbf{px}-Et,  \nonumber
\end{eqnarray}
and introduce the matrices $\widetilde{S}\left( p\right) ,\,\widetilde{S}%
^{T}\left( -p\right) ,\,\widetilde{\mathbf{S}}\left( p\right) ,\,\,\Delta
^{0}$ and $\overline{\Delta }^{0}\,$with the elements $\widetilde{S}%
_{A}^{B}\left( p\right) ,$ \linebreak $\,\widetilde{S}_{A}^{B}\left(
-p\right) ,$ $\,\widetilde{\mathbf{S}}\left( p\right) $ $\left( \Delta
^{0}\right) _{AB}\,$and $\left( \overline{\Delta }^{0}\right) ^{AB}$. Then
we can rewrite the Schwinger-Dyson equation (60) in the matrix form

\begin{equation}
\frac{1}{\widetilde{\mathbf{S}}\left( p\right) }=\frac{1}{\widetilde{S}%
\left( p\right) }+\Delta ^{0}\widetilde{S}^{T}\left( -p\right) \overline{%
\Delta }^{0}.  \label{63}
\end{equation}
For the constant background composite field (55) the field equation (53) has
the explicit form

\begin{eqnarray}
\left( \Delta ^{0}\right) _{AB} &=&-iV_{AB}^{DC}\left\{ \int
d^{4}yS_{C}^{E}\left( x-y\right) \left( \Delta ^{0}\right)
_{EF}S_{D}^{TF}\left( y-x\right) \right.  \nonumber \\
&&\left. -\int d^{4}y\int d^{4}x_{1}\int d^{4}y_{1}S_{C}^{E}\left(
x-y\right) \left( \Delta ^{0}\right) _{EF}S_{D_{1}}^{TF}\left(
y-x_{1}\right) \right.  \nonumber \\
&&\left. .\left( \overline{\Delta }^{0}\right)
^{D_{1}C_{1}}S_{C_{1}}^{E_{1}}\left( x_{1}-y_{1}\right) \left( \Delta
^{0}\right) _{E_{1}F_{1}}S_{D}^{TF_{1}}\left( y_{1}-x\right) \right.
\label{64} \\
&&\left. +\int d^{4}y\int d^{4}x_{1}\int d^{4}y_{1}\int d^{4}x_{2}\int
d^{4}y_{2}S_{C}^{E}\left( x-y\right) \left( \Delta ^{0}\right)
_{EF}S_{D_{1}}^{TF}\left( y-x_{1}\right) \right.  \nonumber \\
&&\left. .\left( \overline{\Delta }^{0}\right)
^{D_{1}C_{1}}S_{C_{1}}^{E_{1}}\left( x_{1}-y_{1}\right) \left( \Delta
^{0}\right) _{E_{1}F_{1}}S_{D_{2}}^{TF_{1}}\left( y_{1}-x_{2}\right) \left( 
\overline{\Delta }^{0}\right) ^{D_{2}C_{2}}\right.  \nonumber \\
&&\left. .S_{C_{2}}^{E_{2}}\left( x_{2}-y_{2}\right) \left( \Delta
^{0}\right) _{E_{2}F_{2}}S_{D}^{TF_{2}}\left( y_{2}-x\right)
-.......\right\} .  \nonumber
\end{eqnarray}
In term of the Fourier transforms of the two-point Green functions this
equation becomes

\begin{eqnarray}
\left( \Delta ^{0}\right) _{AB} &=&-iV_{AB}^{DC}\frac{1}{\left( 2\pi \right)
^{4}}\int d^{4}p\left\{ \widetilde{S}_{C}^{E}\left( p\right) \left( \Delta
^{0}\right) _{EF}\widetilde{S}_{D}^{F}\left( -p\right) \right.  \nonumber \\
&&\left. -\widetilde{S}_{C}^{E}\left( p\right) \left( \Delta ^{0}\right)
_{EF}\widetilde{S}_{D_{1}}^{F}\left( -p\right) \left( \overline{\Delta }%
^{0}\right) ^{D_{1}C_{1}}\widetilde{S}_{C_{1}}^{E_{1}}\left( p\right) \left(
\Delta ^{0}\right) _{E_{1}F_{1}}\widetilde{S}_{D}^{F_{1}}\left( -p\right)
\right.  \nonumber \\
&&\left. +\widetilde{S}_{C}^{E}\left( p\right) \left( \Delta ^{0}\right)
_{EF}\widetilde{S}_{D_{1}}^{F}\left( -p\right) \left( \overline{\Delta }%
^{0}\right) ^{D_{1}C_{1}}\widetilde{S}_{C_{1}}^{E_{1}}\left( p\right) \left(
\Delta ^{0}\right) _{E_{1}F_{1}}\widetilde{S}_{D_{2}}^{F_{1}}\left(
-p\right) \right.  \nonumber \\
&&\left. .\left( \overline{\Delta }^{0}\right) ^{D_{2}C_{2}}\widetilde{S}%
_{C_{2}}^{E_{2}}\left( p\right) \left( \Delta ^{0}\right) _{E_{2}F_{2}}%
\widetilde{S}_{D}^{F_{2}}\left( -p\right) -.......\right\} .  \label{65}
\end{eqnarray}
or in the matrix form

\begin{equation}
\left( \Delta ^{0}\right) _{AB}=-iV_{AB}^{DC}\frac{1}{\left( 2\pi \right)
^{4}}\int d^{4}p\left[ \widetilde{S}\left( p\right) \Delta ^{0}S^{T}\left(
-p\right) \frac{1}{1+\overline{\Delta }^{0}\widetilde{S}\left( p\right)
\Delta ^{0}S^{T}\left( -p\right) }\right] _{CD}  \label{66}
\end{equation}

\section{Relativistic Bethe-Salpeter Equation}

Now we generalize the reasonings presented in the preceding Section to
establish the bound state equations for the two-body systems of relativistic
Dirac fermions with some effective (non-local, in general) $4-$fermion
interaction in the quantum field theory. For the definiteness we consider
again the quark field with some effective direct $4-$quark (non-local, in
general) interaction induced by the dynamical mechanisms in QCD. Introduce
the Fourier transforms $\widetilde{\psi }_{A}\left( p\right) $ and $%
\widetilde{\overline{\psi }}^{A}\left( p\right) $ of the quark field $\psi
_{A}\left( x\right) $ and its conjugate $\overline{\psi }^{A}\left( x\right)
,$

\begin{eqnarray}
\psi _{A}\left( x\right) &=&\frac{1}{\left( 2\pi \right) ^{2}}\int
d^{4}pe^{ipx}\widetilde{\psi }_{A}\left( p\right) ,  \nonumber \\
\overline{\psi }^{A}\left( x\right) &=&\frac{1}{\left( 2\pi \right) ^{2}}%
\int d^{4}pe^{-ipx}\widetilde{\overline{\psi }}^{A}\left( p\right) .
\label{67}
\end{eqnarray}
Then we have

\begin{equation}
S_{0}\left[ \psi ,\overline{\psi }\right] =-\int d^{4}p\widetilde{\overline{%
\psi }}^{A}\left( p\right) \left[ i\left( \widehat{p}\right)
_{A}^{B}+m\delta _{A}^{B}\right] \widetilde{\psi }_{B}\left( p\right)
\label{68}
\end{equation}
and

\begin{equation}
\frac{i}{Z_{0}}\int \left[ D\psi \right] \left[ D\overline{\psi }\right] 
\widetilde{\psi }_{A}\left( p\right) \widetilde{\overline{\psi }}^{B}\left(
q\right) \exp \left\{ iS_{0}\left[ \psi ,\overline{\psi }\right] \right\}
=\delta \left( p-q\right) \widetilde{S}_{A}^{B}\left( p\right)  \label{69}
\end{equation}
The contribution $S_{\text{int}}\left[ \psi ,\overline{\psi }\right] $ of
the $4-$quark interaction to the effective action reproducing the $4-$point
Green function of the quark field will be represented either in the form

\begin{eqnarray}
S_{\text{int}}\left[ \psi ,\overline{\psi }\right] &=&\frac{1}{2}\int
d^{4}p_{1}d^{4}q_{1}\int d^{4}p_{2}d^{4}q_{2}\delta ^{\left( 4\right)
}\left( p_{1}+q_{1}-p_{2}-q_{2}\right)  \label{70}  \\
&&\widetilde{\overline{\psi }}^{A_{1}}\left( p_{1}\right) \widetilde{\psi }%
_{B_{1}}\left( -q_{1}\right) U_{A_{1}B_{2}}^{B_{1}A_{2}}\left(
p_{1},-q_{1};-q_{2},p_{2}\right) \widetilde{\overline{\psi }}^{A_{2}}\left(
-q_{2}\right) \widetilde{\psi }_{B_{2}}\left( p_{2}\right)  \nonumber
\end{eqnarray}
convenient for the study of the mesons as the bound states of the
quark-antiquark pairs, or in another form

\begin{eqnarray}
S_{\text{int}}\left[ \psi ,\overline{\psi }\right] &=&\frac{1}{2}\int
d^{4}p_{1}d^{4}q_{1}\int d^{4}p_{2}d^{4}q_{2}\delta ^{\left( 4\right)
}\left( p_{1}+q_{1}-p_{2}-q_{2}\right)  \label{71}  \\
&&\widetilde{\overline{\psi }}^{A_{1}}\left( p_{1}\right) \widetilde{%
\overline{\psi }}^{B_{1}}\left( q_{1}\right)
V_{B_{1}A_{1}}^{A_{2}B_{2}}\left( q_{1},p_{1};p_{2},q_{2}\right) \widetilde{%
\psi }_{B_{2}}\left( q_{2}\right) \widetilde{\psi }_{A_{2}}\left(
p_{2}\right)  \nonumber
\end{eqnarray}
convenient for the study of the bound states of two quarks - the diquarks.
\noindent The covariant quantities $U_{A_{1}A_{2}}^{B_{1}B_{2}}\left(
p_{1},q_{1};p_{2},q_{2}\right) $ and $V_{B_{1}A_{1}}^{A_{2}B_{2}}\left(
q_{1},p_{1},;p_{2},q_{2}\right) $ may be called the relativistic effective
potentials. They are related each to other

\begin{equation}
V_{B_{1}A_{1}}^{A_{2}B_{2}}\left( q_{1},p_{1};p_{2},q_{2}\right)
=U_{A_{1}B_{1}}^{B_{2}A_{2}}\left( p_{1},p_{2};q_{1},q_{2}\right)  \label{72}
\end{equation}
and must be antisymmetric under the simultaneous interchanges of the upper
or lower spinor indices and the corresponding $4-$momenta of the quark field

\begin{eqnarray}
&&U_{A_{1}B_{2}}^{B_{1}A_{2}}\left( p_{1},-q_{1};-q_{2},p_{2}\right) 
\begin{array}{r}
=
\end{array}
-U_{B_{2}A_{1}}^{A_{2}B_{1}}\left( -q_{2},-q_{1};p_{1},q_{2}\right) 
\begin{array}{r}
=
\end{array}
\nonumber \\
&=&-U_{A_{1}B_{2}}^{A_{2}B_{1}}\left( p_{1},p_{2};-q_{2},-q_{1}\right)
=U_{B_{2}A_{1}}^{A_{2}B_{1}}\left( -q_{2},p_{2};p_{1},-q_{1}\right) , 
\nonumber \\
&&V_{B_{1}A_{1}}^{A_{2}B_{2}}\left( q_{1},p_{1};p_{2},q_{2}\right) 
\begin{array}{r}
=
\end{array}
-V_{A_{1}B_{1}}^{A_{2}B_{2}}\left( p_{1},q_{1};p_{2},q_{2}\right) 
\begin{array}{r}
=
\end{array}
\label{73} \\
&=&-V_{B_{1}A_{1}}^{B_{2}A_{2}}\left( q_{1},p_{1};q_{2},p_{2}\right) 
\begin{array}{r}
=
\end{array}
V_{A_{1}B_{1}}^{B_{2}A_{2}}\left( p_{1},q_{1};q_{2},p_{2}\right) .  \nonumber
\end{eqnarray}

In order to study the mesons we start from the expression (70) of $S_{\text{%
int}}\left[ \psi ,\overline{\psi }\right] $, introduce the composite
bi-local meson field $\Phi _{B}^{A}\left( x,y\right) ,$ whose Fourier
transform is denoted $\widetilde{\Phi }_{B}^{A}\left( p,q\right) $

\begin{equation}
\Phi _{B}^{A}\left( x,y\right) =\frac{1}{\left( 2\pi \right) ^{4}}\int
d^{4}pd^{4}qe^{-i\left( px-qy\right) }\widetilde{\Phi }_{B}^{A}\left(
p,q\right) ,  \label{74}
\end{equation}
and set

\begin{eqnarray}
Z_{0}^{\Phi } &=&\int \left[ D\Phi \right] \exp \left\{ -\frac{i}{2}\int
d^{4}p_{1}d^{4}q_{1}\int d^{4}p_{2}d^{4}q_{2}\delta ^{\left( 4\right)
}\left( p_{1}+q_{1}-p_{2}-q_{2}\right) \right.  \nonumber \\
&&\left. \widetilde{\Phi }_{B_{1}}^{A_{1}}\left( p_{1},-q_{1}\right)
U_{A_{1}B_{2}}^{B_{1}A_{2}}\left( p_{1},-q_{1};-q_{2},p_{2}\right) 
\widetilde{\Phi }_{A_{2}}^{B_{2}}\left( -q_{2},p_{2}\right) \right\}
\label{75}
\end{eqnarray}
Shifting the functional integration variables

\begin{equation}
\widetilde{\Phi }_{B}^{A}\left( p,-q\right) \rightarrow \widetilde{\Phi }%
_{B}^{A}\left( p,-q\right) +\widetilde{\overline{\psi }}^{A}\left( p\right) 
\widetilde{\psi }_{B}\left( -q\right) ,  \label{76}
\end{equation}
we establish the Hubbard-Stratonovich transformation

\begin{eqnarray}
&&\exp \left\{ \frac{i}{2}\int d^{4}p_{1}d^{4}q_{1}\int
d^{4}p_{2}d^{4}q_{2}\delta ^{\left( 4\right) }\left(
p_{1}+q_{1}-p_{2}-q_{2}\right) \right.  \nonumber \\
&&\;\;\;\;\;\left. \widetilde{\overline{\psi }}^{A_{1}}\left( p_{1}\right) 
\widetilde{\psi }_{B_{1}}\left( -q_{1}\right)
U_{A_{1}B_{2}}^{B_{1}A_{2}}\left( p_{1},-q_{1};-q_{2},p_{2}\right) \overline{%
\psi }^{B_{2}}\left( -q_{2}\right) \psi _{A_{2}}\left( p_{2}\right) \right\} 
\begin{array}{r}
=
\end{array}
\nonumber \\
&=&\,\frac{1}{Z_{0}^{\Phi }}\int \left[ D\Phi \right] \exp \left\{ -\frac{i}{%
2}\int d^{4}p_{1}d^{4}q_{1}\int d^{4}p_{2}d^{4}q_{2}\delta ^{\left( 4\right)
}\left( p_{1}+q_{1}-p_{2}-q_{2}\right) \right.  \nonumber \\
&&\;\;\;\;\;\;\left. \widetilde{\Phi }_{B_{1}}^{A_{1}}\left(
p_{1},-q_{1}\right) U_{A_{1}B_{2}}^{B_{1}A_{2}}\left(
p_{1},-q_{1};-q_{2},p_{2}\right) \widetilde{\Phi }_{A_{2}}^{B_{2}}\left(
-q_{2},p_{2}\right) \right\}  \label{77} \\
&&\;\;\;\;\;\;\exp \left\{ -i\int d^{4}pd^{4}q\widetilde{\overline{\psi }}%
^{A}\left( p\right) \widetilde{\psi }_{B}\left( -q\right) \widetilde{\Delta }%
_{A}^{B}\left( p,-q\right) \right\} ,  \nonumber
\end{eqnarray}
where

\begin{equation}
\widetilde{\Delta }_{A}^{B}\left( p,-q\right) =\int d^{4}p^{\prime
}d^{4}q^{\prime }\delta ^{\left( 4\right) }\left( p+q-p^{\prime }-q^{\prime
}\right) U_{AB^{\prime }}^{BA^{\prime }}\left( p,-q;-q^{\prime },p^{\prime
}\right) \widetilde{\Phi }_{A^{\prime }}^{B^{\prime }}\left( -q^{\prime
},p^{\prime }\right) .  \label{78}
\end{equation}
Applying the transformation (77) to the expression in the r.h.s of the
formula (4) with the effective interaction action $S_{\text{int}}\left[ \psi
,\overline{\psi }\right] $ of the form (70) and performing the functional
integration over the fermionic integration variables $\psi _{A}\left(
x\right) $ and $\overline{\psi }_{A}\left( x\right) $ or $\widetilde{\psi }%
_{A}\left( p\right) $ and $\widetilde{\overline{\psi }}^{A}\left( p\right) $%
, we rewrite the functional integral (4) of the interacting quark field in
the form of the functional integral (11) of the composite bi-local meson
field with the effective action

\begin{eqnarray}
S_{\text{eff}}\left[ \Phi \right] &=&-\frac{1}{2}\int
d^{4}p_{1}d^{4}q_{1}\int d^{4}p_{2}d^{4}q_{2}\delta ^{\left( 4\right)
}\left( p_{1}+q_{1}-p_{2}-q_{2}\right)  \label{79} \\
&&\widetilde{\Phi }_{B_{1}}^{A_{1}}\left( p_{1},-q_{1}\right)
U_{A_{1}B_{2}}^{B_{1}A_{2}}\left( p_{1},-q_{1};-q_{2},p_{2}\right) 
\widetilde{\Phi }_{A_{2}}^{B_{2}}\left( -q_{2},p_{2}\right) +W\left[ \Delta
\right] ,  \nonumber
\end{eqnarray}

\noindent where $W\left[ \Delta \right] $ is a functional power series of
the form (15) in the bi-local meson field.

\noindent Calculations give

\begin{equation}
W^{\left( 1\right) }\left[ \Delta \right] =-i\int d^{4}p\widetilde{S}%
_{B}^{A}\left( p\right) \widetilde{\Delta }_{A}^{B}\left( p,-p\right) ,
\label{80}
\end{equation}

\begin{equation}
W^{\left( 2\right) }\left[ \Delta \right] =\frac{i}{2}\int d^{4}p_{1}\int
d^{4}p_{2}\widetilde{S}_{B_{1}}^{A_{1}}\left( p_{1}\right) \widetilde{\Delta 
}_{A_{1}}^{B_{2}}\left( p_{1},-p_{2}\right) \widetilde{S}_{B_{2}}^{A_{2}}%
\left( p_{2}\right) \widetilde{\Delta }_{A_{2}}^{B_{1}}\left(
p_{2},-p_{1}\right) ,  \label{81}
\end{equation}

\begin{eqnarray}
W^{\left( 3\right) }\left[ \Delta \right] &=&-\frac{i}{3}\int d^{4}p_{1}\int
d^{4}p_{2}\int d^{4}p_{3}\widetilde{S}_{B_{1}}^{A_{1}}\left( p_{1}\right) 
\widetilde{\Delta }_{A_{1}}^{B_{2}}\left( p_{1},-p_{2}\right)  \nonumber \\
&&\widetilde{S}_{B_{2}}^{A_{2}}\left( p_{2}\right) \widetilde{\Delta }%
_{A_{2}}^{B_{3}}\left( p_{2},-p_{3}\right) \widetilde{S}_{B_{3}}^{A_{3}}%
\left( p_{3}\right) \widetilde{\Delta }_{A_{3}}^{B_{1}}\left(
p_{3},p_{1}\right) ,  \label{82}
\end{eqnarray}

\noindent etc.\noindent

From the variational principle

\begin{equation}
\frac{\delta S_{\text{eff}}\left[ \Phi \right] }{\delta \widetilde{\Phi }%
_{B}^{A}\left( p,-q\right) }=0  \label{83}
\end{equation}

\noindent it follows the field equation

\begin{equation}
\widetilde{\Delta }_{A}^{B}\left( p,-q\right) =\int d^{4}p^{\prime
}d^{4}q^{\prime }\delta ^{\left( 4\right) }\left( p+q-p^{\prime }-q^{\prime
}\right) U_{AB^{\prime }}^{BA^{\prime }}\left( p,-q;q^{\prime },p^{\prime
}\right) \frac{\delta W\left[ \Delta \right] }{\delta \widetilde{\Delta }%
_{A^{\prime }}^{B^{\prime }}\left( -q^{\prime },p^{\prime }\right) }.
\label{84}
\end{equation}

\noindent If $W^{\left( 1\right) }\left[ \Delta \right] $ vanishes or does
not give the contribution, then in the second order approximation the field
equation (84) becomes a homogeneous linear integral equation for the
bi-local meson field - the relativistic Bethe-Salpeter equation for the
quark-antiquark bound states

\begin{eqnarray}
\widetilde{\Delta }_{A}^{B}\left( p,-q\right) &=&i\int d^{4}p^{\prime
}d^{4}q^{\prime }\delta ^{\left( 4\right) }\left( p+q-p^{\prime }-q^{\prime
}\right)  \nonumber \\
&&U_{AB^{\prime }}^{BA^{\prime }}\left( p,-q;-q^{\prime },p^{\prime }\right)
S_{A^{\prime }}^{C}\left( p^{\prime }\right) \widetilde{\Delta }%
_{C}^{D}\left( p^{\prime },-q^{\prime }\right) S_{D}^{B^{\prime }}\left(
q^{\prime }\right) .  \label{85}
\end{eqnarray}

\noindent In the general case, when $W^{\left( 1\right) }\left[ \Delta
\right] $ is non-vanishing or the contributions of the high-order
functionals $W^{\left( n\right) }\left[ \Delta \right] ,$ $n>1$, are not
negligible, we start to search the solution of the field equation (84) in
the class of the function of the special form

\begin{eqnarray}
\left( \widetilde{\Phi }^{0}\left( p,-q\right) \right) _{B}^{A} &=&\delta
^{\left( 4\right) }\left( p+q\right) \widetilde{\Phi }_{B}^{A}\left(
p\right) ,\qquad \left( \widetilde{\Delta }^{0}\left( p,-q\right) \right)
_{A}^{B}=\delta ^{\left( 4\right) }\left( p+q\right) \widetilde{\Delta }%
_{A}^{B}\left( p\right) , \nonumber \\
\widetilde{\Delta }_{A}^{B}\left( p\right) &=&\int d^{4}p^{\prime
}U_{AA^{\prime }}^{BB^{\prime }}\left( p,-p;p^{\prime },-p^{\prime }\right) 
\widetilde{\Phi }_{B^{\prime }}^{A^{\prime }}\left( -p^{\prime }\right) , 
\label{86}
\end{eqnarray}

\noindent which are the Fourier transforms of the functions $\left( \Phi
^{0}\left( x-y\right) \right) _{B}^{A}$ and $\left( \Delta ^{0}\left(
x-y\right) \right) _{A}^{B}$ depending only on the difference $x-y$ of the
coordinates,

\begin{eqnarray}
\left( \Phi ^{0}\left( x-y\right) \right) _{B}^{A} &=&\frac{1}{\left( 2\pi
\right) ^{4}}\int d^{4}pd^{4}qe^{-i\left( px+qy\right) }\delta ^{\left(
4\right) }\left( p+q\right) \widetilde{\Phi }_{B}^{A}\left( p\right) 
\nonumber \\
&=&\frac{1}{\left( 2\pi \right) ^{4}}\int d^{4}pe^{-ip\left( x+y\right) }%
\widetilde{\Phi }_{B}^{A}\left( p\right) , \\
\left( \Delta ^{0}\left( x-y\right) \right) _{A}^{B} &=&\frac{1}{\left( 2\pi
\right) ^{4}}\int d^{4}pd^{4}qe^{-i\left( px+qy\right) }\delta ^{\left(
4\right) }\left( p+q\right) \widetilde{\Delta }_{A}^{B}\left( p\right) 
\nonumber \\
&&\frac{1}{\left( 2\pi \right) ^{4}}\int d^{4}pe^{-ip\left( x+y\right) }%
\widetilde{\Delta }_{A}^{B}\left( p\right) ,  \nonumber
\end{eqnarray}

\noindent and then to consider the quantum fluctuations of the bi-local
meson field around the background one of the form (87)

\begin{center}
\begin{eqnarray}
\widetilde{\Phi }_{B}^{A}\left( p,-q\right) &=&\delta ^{\left( 4\right)
}\left( p-q\right) \widetilde{\Phi }_{B}^{A}\left( p\right) +\widetilde{%
\varphi }_{B}^{A}\left( p,-q\right) ,  \nonumber \\
\widetilde{\Delta }_{A}^{B}\left( p,-q\right) &=&\delta ^{\left( 4\right)
}\left( p-q\right) \widetilde{\Delta }_{A}^{B}\left( p\right) +\widetilde{%
\xi }_{A}^{B}\left( p,-q\right) ,  \label{88}
\end{eqnarray}

\begin{equation}
\widetilde{\xi }_{A}^{B}\left( p,-q\right) =\int d^{4}p^{\prime
}d^{4}q^{\prime }\delta ^{\left( 4\right) }\left( p+q-p^{\prime }-q^{\prime
}\right) U_{AB^{\prime }}^{BA^{\prime }}\left( p,-q;-q^{\prime },p^{\prime
}\right) \widetilde{\xi }_{A^{\prime }}^{B^{\prime }}\left( -q^{\prime
},p^{\prime }\right) .  \label{89}
\end{equation}
\end{center}

\noindent For the background meson field (87) the equation (84) becomes

\begin{equation}
\widetilde{\Delta }_{A}^{B}\left( p\right) =-i\int d^{4}qU_{AC}^{BD}\left(
p,-p;-q,q\right) \widetilde{\mathbf{S}}_{D}^{C}\left( q\right) ,  \label{90}
\end{equation}
where $\widetilde{\mathbf{S}}_{A}^{B}\left( p\right) $ is the Fourier
transform of the two-point Green function of the quarks interacting with the
meson field (86) :

\begin{eqnarray}
\widetilde{\mathbf{S}}_{A}^{B}\left( p\right) &=&\widetilde{S}_{A}^{B}\left(
p\right) -\widetilde{S}_{A}^{A_{1}}\left( p\right) \widetilde{\Delta }%
_{A_{1}}^{B_{1}}\left( p\right) \widetilde{S}_{B_{1}}^{B}\left( p\right) 
\nonumber \\
&&+\widetilde{S}_{A}^{A_{1}}\left( p\right) \widetilde{\Delta }%
_{A_{1}}^{B_{1}}\left( p\right) \widetilde{S}_{B_{1}}^{A_{2}}\left( p\right) 
\widetilde{\Delta }_{A_{2}}^{B_{2}}\left( p\right) \widetilde{S}%
_{B_{2}}^{B}\left( p\right) -.....\mathbf{.}  \label{91}
\end{eqnarray}

\noindent It satisfies the Schwinger-Dyson equation

\begin{equation}
\widetilde{\mathbf{S}}_{A}^{B}\left( p\right) =\widetilde{S}_{A}^{B}\left(
p\right) -\widetilde{S}_{A}^{C}\left( p\right) \widetilde{\Delta }%
_{C}^{D}\left( p\right) \widetilde{\mathbf{S}}_{D}^{B}\left( p\right) .
\label{92}
\end{equation}

\noindent In the second order with respect to the quantum fluctuations the
effective action (79) equals

\begin{eqnarray}
S_{\text{eff}}\left[ \Phi \right] &=&S_{\text{eff}}\left[ \Phi ^{0}\right] -%
\frac{1}{2}\int d^{4}p_{1}d^{4}q_{1}\int d^{4}p_{2}d^{4}q_{2}\delta ^{\left(
4\right) }\left( p_{1}+p_{2}-q_{1}-q_{2}\right)  \nonumber \\
&&\widetilde{\varphi }_{B_{1}}^{A_{1}}\left( p_{1},-q_{1}\right)
U_{A_{1}B_{2}}^{B_{1}A_{2}}\left( p_{1},-q_{1};-q_{2},p_{2}\right) 
\widetilde{\varphi }_{A_{2}}^{B_{2}}\left( -q_{2},p_{2}\right) +  \nonumber
\\
&&+\frac{1}{2}\int d^{4}p_{1}d^{4}q_{1}\int d^{4}p_{2}d^{4}q_{2}\,.
\label{93} \\
&&\,\,\,\,\,\widetilde{\xi }_{B_{1}}^{A_{1}}\left( p_{1},-q_{1}\right) \frac{%
\delta ^{2}W\left[ \Delta \right] }{\delta \widetilde{\Delta }%
_{A_{1}}^{B_{1}}\left( p_{1},-q_{1}\right) \delta \widetilde{\Delta }%
_{A_{2}}^{B_{2}}\left( p_{2},-q_{2}\right) }|_{\widetilde{\Delta }%
_{A}^{B}=\left( \widetilde{\Delta }^{0}\right) _{A}^{B}}\widetilde{\xi }%
_{B_{2}}^{A_{2}}\left( p_{2},-q_{2}\right) .  \nonumber
\end{eqnarray}

\noindent By summing up the infinite series

\[
\sum\limits_{n=1}^{\infty }\frac{\delta ^{2}W^{\left( n\right) }\left[
\Delta \right] }{\delta \widetilde{\Delta }_{A_{1}}^{B_{1}}\left(
p_{1},-q_{1}\right) \delta \widetilde{\Delta }_{A_{2}}^{B_{2}}\left(
p_{2},-q_{2}\right) }|_{\widetilde{\Delta }_{A}^{B}=\left( \widetilde{\Delta 
}^{0}\right) _{A}^{B}} 
\]

\noindent we obtain

\begin{equation}
\frac{\delta ^{2}W\left[ \Delta \right] }{\delta \widetilde{\Delta }%
_{A_{1}}^{B_{1}}\left( p_{1},-q_{1}\right) \delta \widetilde{\Delta }%
_{A_{2}}^{B_{2}}\left( p_{2},-q_{2}\right) }|_{\widetilde{\Delta }%
_{A}^{B}=\left( \widetilde{\Delta }^{0}\right) _{A}^{B}}=i\delta ^{\left(
4\right) }\left( p_{1}+q_{2}\right) \delta ^{\left( 4\right) }\left(
p_{2}+q_{1}\right) \widetilde{\mathbf{S}}_{A_{1}}^{B_{2}}\left( p_{1}\right) 
\widetilde{\mathbf{S}}_{A_{2}}^{B_{1}}\left( p_{2}\right) .\qquad  \label{94}
\end{equation}

\noindent Then from the formula (93) for the effective action and the
variational principle it follows the relativistic Bethe-Salpeter equation
containing the two-point Green function of the quarks interacting with the
background meson field

\begin{eqnarray}
\,\widetilde{\xi }_{A}^{B}\left( p,-q\right) &=&\int d^{4}p^{\prime
}d^{4}q^{\prime }\delta ^{\left( 4\right) }\left( p+q-p^{\prime }-q^{\prime
}\right)  \nonumber \\
&&U_{AB^{\prime }}^{BA^{\prime }}\left( p,-q;-q^{\prime },p^{\prime }\right) 
\widetilde{\mathbf{S}}_{A^{\prime }}^{C}\left( p^{\prime }\right) \widetilde{%
\xi }_{C}^{D}\left( p^{\prime },-q^{\prime }\right) \widetilde{\mathbf{S}}%
_{D}^{B^{\prime }}\left( q^{\prime }\right) .  \label{95}
\end{eqnarray}
In the special case when the background meson field vanishes or its
contribution is neglected then the equation (95) reduces to the equation (85
)

In order to study the formation of the diquarks we start from the expression
(71) of $S_{\text{int}}\left[ \psi ,\overline{\psi }\right] $, introduce the
composite bi-local bi-spinor bosonic field $\Phi _{AB}\left( x,y\right) $
and its conjugate $\overline{\Phi }^{AB}\left( x,y\right) $ as well as their
Fourier transforms $\widetilde{\Phi }_{AB}\left( p,q\right) $ and $%
\widetilde{\overline{\Phi }}^{AB}\left( p,q\right) $

\begin{eqnarray}
\Phi _{AB}\left( x,y\right) &=&\frac{1}{\left( 2\pi \right) ^{4}}\int
d^{4}pd^{4}qe^{i\left( px+qy\right) }\widetilde{\Phi }_{AB}\left( p,q\right)
,  \nonumber \\
\overline{\Phi }^{AB}\left( x,y\right) &=&\frac{1}{\left( 2\pi \right) ^{4}}%
\int d^{4}pd^{4}qe^{-i\left( px+qy\right) }\widetilde{\overline{\Phi }}%
^{AB}\left( p,q\right) ,  \label{96}
\end{eqnarray}
and set

\begin{eqnarray}
Z_{0}^{\Phi ,\overline{\Phi }} &=&\int \left[ D\Phi \right] \left[ D%
\overline{\Phi }\right] \exp \left\{ -\frac{i}{2}\int d^{4}p_{1}\int
d^{4}q_{1}\int d^{4}p_{2}\int d^{4}q_{2}\delta ^{4}\left(
p_{1}+q_{1}-p_{2}-q_{2}\right) \right.  \nonumber \\
&&\,\,\,\,\,\,\,\,\,\,\,\,\,\,\,\,\,\,\,\,\,\,\,\,\,\,\,\,\,\,\,\,\,\,\,\,\,%
\,\,\,\,\,\,\left. \widetilde{\overline{\Phi }}^{A_{1}B_{1}}\left(
p_{1},q_{1}\right) V_{B_{1}A_{1}}^{A_{2}B_{2}}\left(
q_{1},p_{1};p_{2},q_{2}\right) \widetilde{\Phi }_{B_{2}A_{2}}\left(
q_{2},p_{2}\right) \right\}  \label{97}
\end{eqnarray}

\noindent Shifting the functional integration variables

\begin{eqnarray}
\widetilde{\Phi }_{AB}\left( p,q\right) &\rightarrow &\widetilde{\Phi }%
_{AB}\left( p,q\right) +\widetilde{\psi }_{A}\left( p\right) \widetilde{\psi 
}_{B}\left( q\right) ,  \nonumber \\
\widetilde{\overline{\Phi }}^{AB}\left( p,q\right) &\rightarrow &\widetilde{%
\overline{\Phi }}^{AB}\left( p,q\right) +\widetilde{\overline{\psi }}%
^{A}\left( p\right) \widetilde{\overline{\psi }}^{B}\left( q\right) ,
\label{98}
\end{eqnarray}
we establish the Hubbard-Stratonovich transformation

\begin{eqnarray}
&&\exp \left\{ \frac{i}{2}\int d^{4}p_{1}\int d^{4}q_{1}\int d^{4}p_{2}\int
d^{4}q_{2}\delta ^{4}\left( p_{1}+q_{1}-p_{2}-q_{2}\right) \right.  \nonumber
\\
&&\,\,\,\,\,\,\,\,\,\,\,\,\,\,\,\left. \widetilde{\overline{\psi }}%
^{A_{1}}\left( p_{1}\right) \widetilde{\overline{\psi }}^{B_{1}}\left(
q_{1}\right) V_{B_{1}A_{1}}^{A_{2}B_{2}}\left(
q_{1},p_{1};p_{2},q_{2}\right) \widetilde{\psi }_{B_{2}}\left( q_{2}\right) 
\widetilde{\psi }_{A_{2}}\left( p_{2}\right) \right\}  \nonumber \\
&=&\frac{1}{Z_{0}^{\Phi ,\overline{\Phi }}}\int \left[ D\Phi \right] \left[ D%
\overline{\Phi }\right] \exp \left\{ -\frac{i}{2}\int d^{4}p_{1}\int
d^{4}q_{1}\int d^{4}p_{2}\int d^{4}q_{2}\delta ^{\left( 4\right) }\left(
p_{1}+q_{1}-p_{2}-q_{2}\right) \right.  \nonumber \\
&&\,\,\,\,\,\,\,\,\,\,\,\,\,\,\,\,\,\,\,\,\left. \widetilde{\overline{\Phi }}%
^{A_{1}B_{1}}\left( p_{1},q_{1}\right) V_{B_{1}A_{1}}^{A_{2}B_{2}}\left(
q_{1},p_{1};p_{2},q_{2}\right) \widetilde{\Phi }_{B_{2}A_{2}}\left(
q_{2},p_{2}\right) \right\}  \label{99} \\
&&.\exp \left\{ -\frac{i}{2}\int d^{4}p\int d^{4}q\widetilde{\overline{\psi }%
}^{A}\left( p\right) \widetilde{\overline{\psi }}^{B}\left( q\right) 
\widetilde{\Delta }_{BA}\left( q,p\right) +\widetilde{\overline{\Delta }}%
^{AB}\left( p,q\right) \widetilde{\psi }_{B}\left( q\right) \widetilde{\psi }%
_{A}\left( p\right) \right\} ,  \nonumber
\end{eqnarray}

\begin{eqnarray}
\widetilde{\Delta }_{BA}\left( q,p\right) &=&\int d^{4}p^{\prime }\int
d^{4}q^{\prime }\delta ^{\left( 4\right) }\left( p^{\prime }+q^{\prime
}-p-q\right) V_{BA}^{A^{\prime }B^{\prime }}\left( q,p;p^{\prime },q^{\prime
}\right) \widetilde{\Phi }_{B^{\prime }A^{\prime }}\left( q^{\prime
},p^{\prime }\right) ,  \nonumber \\
\widetilde{\overline{\Delta }}^{AB}\left( p,q\right) &=&\int d^{4}p^{\prime
}\int d^{4}q^{\prime }\delta ^{\left( 4\right) }\left( p^{\prime }+q^{\prime
}-p-q\right) \widetilde{\overline{\Phi }}^{A^{\prime }B^{\prime }}\left(
p^{\prime },q^{\prime }\right) V_{B^{\prime }A^{\prime }}^{AB}\left(
q^{\prime },p^{\prime };p,q\right) . \hskip 1cm  \label{100}
\end{eqnarray}
The bi-local bi-spinor fields $\widetilde{\Phi }_{AB}\left( p,q\right) $ and 
$\widetilde{\overline{\Phi }}^{AB}\left( p,q\right) \,$ are antisymmetric
under the simultaneous permutations of the coordinates and the bi-spinor
indices

\begin{eqnarray}
\widetilde{\Phi }_{BA}\left( q,p\right) &=&-\widetilde{\Phi }_{AB}\left(
p,q\right) ,\qquad \widetilde{\Delta }_{BA}\left( q,p\right) =-\widetilde{%
\Delta }_{AB}\left( p,q\right) ,  \nonumber \\
\widetilde{\overline{\Phi }}^{BA}\left( q,p\right) &=&-\widetilde{\overline{%
\Phi }}^{AB}\left( p,q\right) ,\qquad \widetilde{\overline{\Delta }}%
^{BA}\left( q,p\right) =-\widetilde{\overline{\Delta }}^{AB}\left(
p,q\right) .  \label{101}
\end{eqnarray}
Substituting the expression in the r.h.s. of the relation (99) into the
functional integral (4) and performing the functional integration over the
fermionic integration variables $\psi _{A}\left( x\right) $ and $\overline{%
\psi }^{A}\left( x\right) $ or $\widetilde{\psi }_{A}\left( p\right) $ and $%
\widetilde{\overline{\psi }}^{A}\left( p\right) ,$ we rewrite the functional
integral (4) of the interacting quark field in the form of the functional
integral (43) of the composite bi-local bi-spinor bosonic diquark field with
the effective action

\begin{eqnarray}
S_{\text{eff}}\left[ \Phi ,\overline{\Phi }\right] &=&-\frac{1}{2}\int
d^{4}p_{1}\int d^{4}q_{1}\int d^{4}p_{2}\int d^{4}q_{2}\delta ^{4}\left(
p_{1}+q_{1}-p_{2}-q_{2}\right)  \label{102} \\
&&\,\,\,\,\,\widetilde{\,\overline{\Phi }}^{A_{1}B_{1}}\left(
p_{1},q_{1}\right) V_{B_{1}A_{1}}^{A_{2}B_{2}}\left(
q_{1},p_{1};p_{2},q_{2}\right) \widetilde{\Phi }_{B_{2}A_{2}}\left(
q_{2},p_{2}\right) +W\left[ \Delta ,\overline{\Delta }\right] ,  \nonumber
\end{eqnarray}
where $W\left[ \Delta ,\overline{\Delta }\right] $ is a functional power
series of the form (46) in the fields $\Delta _{AB}\left( x,y\right) $ and $%
\overline{\Delta }^{AB}\left( x,y\right) $ whose Fourier transforms are $%
\widetilde{\Delta }_{AB}\left( p,q\right) $ and $\widetilde{\overline{\Delta 
}}^{AB}\left( p,q\right) $ resp.,

\begin{eqnarray}
\Delta _{AB}\left( x,y\right) &=&\frac{1}{\left( 2\pi \right) ^{4}}\int
d^{4}pd^{4}qe^{i\left( px+qy\right) }\widetilde{\Delta }_{AB}\left(
p,q\right) ,  \nonumber \\
\overline{\Delta }^{AB}\left( x,y\right) &=&\frac{1}{\left( 2\pi \right) ^{4}%
}\int d^{4}pd^{4}qe^{-i\left( px+qy\right) }\widetilde{\overline{\Delta }}%
^{AB}\left( p,q\right) .
\end{eqnarray}
Calculations give

\begin{equation}
W^{\left( 2\right) }\left[ \Delta ,\overline{\Delta }\right] =-\frac{i}{2}%
\int d^{4}p\int d^{4}q\widetilde{\overline{\Delta }}^{AB}\left( p,q\right)
S_{B}^{D}\left( q\right) S_{A}^{C}\left( p\right) \widetilde{\Delta }%
_{DC}\left( q,p\right)  \label{104}
\end{equation}

\begin{eqnarray}
W^{\left( 4\right) }\left[ \Delta ,\overline{\Delta }\right] &=&\frac{i}{4}%
\int d^{4}p_{1}\int d^{4}q_{1}\int d^{4}p_{2}\int d^{4}q_{2}\widetilde{%
\overline{\Delta }}^{A_{1}B_{1}}\left( p_{1},q_{1}\right)
S_{B_{1}}^{D_{1}}\left( q_{1}\right) \widetilde{\Delta }_{D_{1}C_{1}}\left(
q_{1},p_{1}\right)  \nonumber \\
&&S_{A_{2}}^{C_{1}}\left( p_{2}\right) \widetilde{\overline{\Delta }}%
^{A_{2}B_{2}}\left( p_{2},q_{2}\right) S_{B_{2}}^{D_{2}}\left( q_{2}\right) 
\widetilde{\Delta }_{D_{2}C_{2}}\left( q_{2},p_{2}\right)
S_{A_{1}}^{C_{2}}\left( p_{1}\right) ,\qquad
\end{eqnarray}
\begin{eqnarray}
W^{\left( 6\right) }\left[ \Delta ,\overline{\Delta }\right] &=&\frac{i}{6}%
\int d^{4}p_{1}\int d^{4}q_{1}\int ...\int d^{4}p_{3}\int d^{4}q_{3}%
\widetilde{\overline{\Delta }}^{A_{1}B_{1}}\left( p_{1},q_{1}\right)
S_{B_{1}}^{D_{1}}\left( q_{1}\right) \widetilde{\Delta }_{D_{1}C_{1}}\left(
q_{1},p_{1}\right)  \nonumber \\
&&S_{A_{2}}^{C_{1}}\left( p_{2}\right) \widetilde{\overline{\Delta }}%
^{A_{2}B_{2}}\left( p_{2},q_{2}\right) .....S_{B_{3}}^{D_{3}}\left(
q_{3}\right) \widetilde{\Delta }_{D_{3}C_{3}}\left( q_{3},p_{3}\right)
S_{A_{1}}^{C_{3}}\left( p_{1}\right) ,  \label{106}
\end{eqnarray}

\noindent etc.

From the variational principle

\begin{equation}
\frac{\delta S_{\text{eff}}\left[ \Phi ,\overline{\Phi }\right] }{\widetilde{%
\delta \overline{\Phi }}^{AC}\left( x,y\right) }=0  \label{107}
\end{equation}
it follows the field equation

\begin{equation}
\frac{1}{2}\widetilde{\Delta }_{CA}\left( q,p\right) =\int d^{4}p^{\prime
}\int d^{4}q^{\prime }\delta ^{\left( 4\right) }\left( p+q-p^{\prime
}-q^{\prime }\right) V_{CA}^{BD}\left( p,q;p^{\prime },q^{\prime }\right) 
\frac{\delta W\left[ \Delta ,\overline{\Delta }\right] }{\delta \widetilde{%
\overline{\Delta }}^{BD}\left( p^{\prime },q^{\prime }\right) }.  \label{108}
\end{equation}
In the second order approximation with respect to the composite diquark
field we have the linear integral equation

\begin{eqnarray}
\widetilde{\Delta }_{CA}\left( q,p\right) &=&-i\int d^{4}p^{\prime }\int
d^{4}q^{\prime }\delta ^{\left( 4\right) }\left( p+q-p^{\prime }-q^{\prime
}\right)  \nonumber \\
&&V_{CA}^{BD}\left( p,q;p^{\prime },q^{\prime }\right) S_{B}^{E}\left(
p^{\prime }\right) S_{D}^{F}\left( q^{\prime }\right) \widetilde{\Delta }%
_{FE}\left( q^{\prime },p^{\prime }\right) ,\qquad  \label{109}
\end{eqnarray}
which is the relativistic Bethe-Salpeter equation for the diquarks. In the
general case, when the contributions of the high order functionals $%
W^{\left( 2n\right) }\left[ \Delta ,\overline{\Delta }\right] ,\,n>1$,
cannot be neglected, we search the solution $\left( \Phi ^{0}\left(
y-x\right) \right) _{DB},\left( \,\overline{\Phi }^{0}\left( x-y\right)
\right) ^{BD}$ or $\left( \Delta ^{0}\left( y-x\right) \right)
_{CA},\,
\newline 
\left( \overline{\Delta }^{0}\left( x-y\right) \right) ^{AC}$ of the
field equation in the class of the functions depending only on the
difference $x-y$ of the coordinates and study the quantum fluctuations
around this background field:

\begin{eqnarray}
\Phi _{DB}\left( y,x\right) &=&\left( \Phi ^{0}\left( y-x\right) \right)
_{DB}+\varphi _{DB}\left( y,x\right) ,  \nonumber \\
\,\overline{\Phi }^{BD}\left( x,y\right) &=&\left( \overline{\Phi }%
^{0}\left( x-y\right) \right) ^{BD}+\overline{\varphi }^{BD}\left(
x,y\right) ,  \nonumber \\
\Delta _{CA}\left( y,x\right) &=&\left( \Delta ^{0}\left( y-x\right) \right)
_{CA}+\xi _{CA}\left( y,x\right) ,  \nonumber \\
\overline{\Delta }^{AC}\left( x,y\right) &=&\left( \overline{\Delta }%
^{0}\left( x-y\right) \right) ^{AC}+\overline{\xi }^{AC}\left( x,y\right) .
\label{110}
\end{eqnarray}
The Fourier transforms of the background field and its conjugate have
special form

\begin{eqnarray}
\left( \widetilde{\Phi }^{0}\left( q,p\right) \right) _{DB} &=&\delta
^{\left( 4\right) }\left( p+q\right) \widetilde{\Phi }_{DB}\left( p\right)
,\,\,\left( \widetilde{\overline{\Phi }}^{0}\left( p,q\right) \right)
^{BD}=\delta ^{\left( 4\right) }\left( p+q\right) \widetilde{\overline{\Phi }%
}^{BD}\left( p\right) ,  \nonumber \\
\left( \widetilde{\Delta }^{0}\left( q,p\right) \right) _{CA} &=&\delta
^{\left( 4\right) }\left( p+q\right) \widetilde{\Delta }_{CA}\left( p\right)
,\,\,\,\,\left( \widetilde{\overline{\Delta }}^{0}\left( p,q\right) \right)
^{AC}=\delta ^{\left( 4\right) }\left( p+q\right) \widetilde{\overline{%
\Delta }}^{AC}\left( p\right) ,\hskip 1cm  \label{111}
\end{eqnarray}
Denote $\widetilde{\varphi }_{DB}\left( q,p\right) ,\,\widetilde{\overline{%
\varphi }}^{BD}\left( p,q\right) $ and $\widetilde{\xi }_{CA}\left(
q,p\right) ,\,\,\widetilde{\overline{\xi }}^{AC}\left( p,q\right) $ the
Fourier transforms of the fields $\varphi _{DB}\left( y,x\right) ,\,\,%
\overline{\varphi }^{BD}\left( x,y\right) $ and $\xi _{CA}\left( y,x\right)
,\,\,\overline{\xi }^{AC}\left( x,y\right) $, resp.. In the second order
with respect to the quantum fluctuations the effective action equals
\begin{eqnarray}
S_{\text{eff}}\left[ \Phi ,\overline{\Phi }\right] &\approx &S_{\text{eff}%
}\left[ \Phi ^{0},\overline{\Phi }^{0}\right] -\frac{1}{2}\int
d^{4}p_{1}\int d^{4}q_{1}\int d^{4}p_{2}\int d^{4}q_{2}\delta ^{\left(
4\right) }\left( p_{1}+q_{1}-p_{2}-q_{2}\right)  \nonumber \\
&&\widetilde{\overline{\varphi }}^{A_{1}C_{1}}\left( p_{1},q_{1}\right)
V_{C_{1}A_{1}}^{A_{2}C_{2}}\left( p_{1},q_{1};p_{2},q_{2}\right) \widetilde{%
\varphi }_{C_{2}A_{2}}\left( q_{2},p_{2}\right)  \nonumber \\
&&+\int d^{4}p_{1}\int d^{4}q_{1}\int d^{4}p_{2}\int d^{4}q_{2}  \label{112}
\\
&&\widetilde{\overline{\xi }}^{A_{1}C_{1}}\left( p,q\right) \frac{\delta
^{2}W\left[ \Delta ^{0},\overline{\Delta }^{0}\right] }{\delta \widetilde{%
\overline{\Delta }}^{A_{1}C_{1}}\left( p_{1},q_{1}\right) \delta \Delta
_{C_{2}A_{2}}\left( q_{2},p_{2}\right) } \vert_{\Delta _{CA}=\left( \Delta
^{0}\right) _{CA}, \overline{\Delta }^{AC}=\left( \overline{\Delta }%
^{0}\right) ^{AC}} \widetilde{\xi }_{C_{2}A_{2}}\left(
q_{2},p_{2}\right) .  \nonumber
\end{eqnarray}
Summing up the infinite series

\[
\sum\limits_{n=1}^{\infty }\frac{\delta ^{2}W\left[ \Delta ^{0},\overline{%
\Delta }^{0}\right] }{\delta \widetilde{\overline{\Delta }}%
^{A_{1}C_{1}}\left( p_{1},q_{1}\right) \delta \Delta _{C_{2}A_{2}}\left(
q_{2},p_{2}\right) }\vert_{{\Delta _{CA}=\left( \Delta ^{0}\right) _{CA}  \\ 
\overline{\Delta }^{AC}=\left( \overline{\Delta }^{0}\right) ^{AC}}}  , 
\]
we obtain

\begin{eqnarray}
&&\frac{\delta ^{2}W\left[ \Delta ^{0},\overline{\Delta }^{0}\right] }{%
\delta \widetilde{\overline{\Delta }}^{A_{1}C_{1}}\left( p_{1},q_{1}\right)
\delta \Delta _{C_{2}A_{2}}\left( q_{2},p_{2}\right) }\vert_{{\Delta
_{CA}=\left( \Delta ^{0}\right) _{CA},\overline{\Delta }^{AC}=\left( 
\overline{\Delta }^{0}\right) ^{AC}}}=  \nonumber  \\
&=&-\frac{i}{2}\delta ^{\left( 4\right) }\left( p_{1}-p_{2}\right) \delta
^{\left( 4\right) }\left( q_{1}-q_{2}\right) \widetilde{\mathbf{S}}%
_{C_{1}}^{C_{2}}\left( q_{1}\right) \widetilde{\mathbf{S}}%
_{A_{1}}^{A_{2}}\left( p_{1}\right) ,  \label{113}
\end{eqnarray}
where

\begin{eqnarray}
\widetilde{\mathbf{S}}_{A}^{C}\left( p\right) &=&\widetilde{S}_{A}^{C}\left(
p\right) -\widetilde{S}_{A}^{C_{1}}\left( p\right) \widetilde{\Delta }%
_{C_{1}A_{1}}\left( p\right) \,\,S_{A_{2}}^{A_{1}}\left( -p\right) 
\widetilde{\overline{\Delta }}^{A_{2}C_{2}}\left( p\right) \widetilde{S}%
_{C_{2}}^{C}\left( p\right)  \nonumber \\
&&+\widetilde{S}_{A}^{C_{1}}\left( p\right) \widetilde{\Delta }%
_{C_{1}A_{1}}\left( p\right) S_{A_{2}}^{A_{1}}\left( -p\right) \widetilde{%
\overline{\Delta }}^{A_{2}C_{2}}\left( p\right) \widetilde{S}%
_{C_{2}}^{C_{3}}\left( p\right) \widetilde{\Delta }_{C_{3}A_{3}}\left(
p\right)  \nonumber \\
&&.S_{A_{4}}^{A_{3}}\left( -p\right) \widetilde{\overline{\Delta }}%
^{A_{4}C_{4}}\left( p\right) S_{C_{4}}^{C}\left( p\right) -.....  \label{114}
\end{eqnarray}
is the two-point Green function of the quark field in the presence of the
background diquark one and satisfies the Schwinger-Dyson equation

\begin{equation}
\widetilde{\mathbf{S}}_{A}^{C}\left( p\right) =\widetilde{S}_{A}^{C}\left(
p\right) -\widetilde{S}_{A}^{C_{1}}\left( p\right) \widetilde{\Delta }%
_{C_{1}A_{1}}\left( p\right) \,\,S_{A_{2}}^{A_{1}}\left( -p\right) 
\widetilde{\overline{\Delta }}^{A_{2}C_{2}}\left( p\right) \widetilde{%
\mathbf{S}}_{C_{2}}^{C}\left( p\right) ,  \label{115}
\end{equation}
or in the matrix form

\begin{equation}
\frac{1}{\widetilde{\mathbf{S}}\left( p\right) }=\frac{1}{\widetilde{S}%
\left( p\right) }+\widetilde{\Delta }\left( p\right) S^{\text{T}}\left(
-p\right) \widetilde{\overline{\Delta }}\left( p\right) ,  \label{116}
\end{equation}
where $\widetilde{\Delta }\left( p\right) $ and $\widetilde{\overline{\Delta 
}}\left( p\right) $ are the matrices with the elements $\widetilde{\Delta }%
_{CA}\left( p\right) $ and $\,\widetilde{\overline{\Delta }}^{AC}\left(
p\right) .$

\noindent From the expression (112) of the effective action and the relation
(113) it follows the relativistic Bethe-Salpeter equation

\begin{eqnarray}
\widetilde{\xi }_{CA}\left( q,p\right) &=&-i\int d^{4}p^{\prime }\int
d^{4}q^{\prime }\delta ^{\left( 4\right) }\left( p+q-p^{\prime }-q^{\prime
}\right) \\
&&V_{CA}^{BD}\left( p,q;p^{\prime },q^{\prime }\right) \widetilde{\mathbf{S}}%
_{B}^{E}\left( p^{\prime }\right) \widetilde{\mathbf{S}}_{D}^{F}\left(
q^{\prime }\right) \widetilde{\xi }_{FE}\left( q^{\prime },p^{\prime }\right)
\nonumber
\end{eqnarray}
containing the two-point Green function of the quark field interacting with
the background diquark one. If the later vanishes, then the equation (117)
reduces to the equation (109).

The relativistic effective potential $U_{A_{1}B_{2}}^{B_{1}A_{2}}\left(
p_{1},-q_{1};-q_{2},p_{2}\right) $ and $V_{B_{1}A_{1}}^{A_{2}B_{2}}\left(
q_{1},p_{1};-p_{2},q_{2}\right) $ in the bound state equation (85), (95) and
(109), (117) are determined by the dynamical mechanism of the effective $4-$%
quark interaction in QCD. In the case of the instanton induced $4-$quark
coupling they are expressed in terms of the form-factors calculated in the
work of Rapp, Sch\"{a}fer, Shuryak and Velkovsky$^{\left[ 31\right] }$. For
the effective non-local $4-$quark interaction generated by the one-gluon
exchange mechanism we have

\begin{eqnarray}
U_{A_{1}B_{2}}^{B_{1}A_{2}}\left( p_{1},-q_{1};-q_{2},p_{2}\right) &=&4\pi
\sum\limits_{I}\left\{ \frac{\alpha \left[ \left( p_{1}+q_{1}\right)
^{2}\right] }{\left( p_{1}+q_{1}\right) ^{2}}\left( \gamma _{\mu }\otimes
\lambda _{I}\right) _{A_{1}}^{B_{1}}\left( \gamma _{\mu }\otimes \lambda
_{I}\right) _{B_{2}}^{A_{2}}\right.  \nonumber \\
&&\left. -\frac{\alpha \left[ \left( p_{1}-p_{2}\right) ^{2}\right] }{\left(
p_{1}-p_{2}\right) ^{2}}\left( \gamma _{\mu }\otimes \lambda _{I}\right)
_{A_{1}}^{A_{2}}\left( \gamma _{\mu }\otimes \lambda _{I}\right)
_{B_{2}}^{B_{1}}\right\} ,  \label{118}
\end{eqnarray}
where $\alpha \left( k\right) ^{2}$ is the running quark-gluon coupling
constant in QCD, $\lambda _{I}$ are the Gell-Mann matrices of the color
symmetry group $SU\left( N_{C}\right) $; the flavor indices play no role and
are omitted.. In reality there do exist the contribution of both above
mentioned dynamical mechanisms of the effective $4$-quark interactions, and
they must lie taken into account simultaneously. Note that in the special
case of the direct $4$-quark coupling with the interaction Lagrangian (1) or
(2) the Bethe-Salpeter equation (5) and (117) reduce to the
Nambu-Jona-Lasinio equations (38) and (61) resp., This means that the
Nambu-Jona-Lasinio equation is a particular form of the Bethe-Salpeter
equation.

\section{ Bethe-Salpeter Equation for Bipolaritons.}

Now we apply the method presented in two preceding Sections to the study of
the bound states of two quasiparticles in the condensed matters. As a
typical example of the system of two quasiparticles with the complicated
dispersion curves and the momentum dependent interaction potential energies
we consider the bipolariton-the bound states of two excitonic polaritons in
semiconductors. The excitonic polaritons are the elementary excitations
whose formation is the consequence of the quantum mutual transition between
the photons and the excitons-the exciton-photon mixing. Denote $\gamma
_{\sigma }\left( \mathbf{p}\right) $ and $\gamma _{\sigma }^{+}\left( 
\mathbf{p}\right) $ the destruction and creation operators of the photon
with the momentum\textbf{\ }$\mathbf{p}$ and the energy $\omega (\mathbf{p})$
in the polarization state labeled by the index $a=1,2$ , where

\[
\omega (\mathbf{p})=\varepsilon p, 
\]
$\varepsilon $\ being the background dielectric constant of the
semiconductor. Due to the electromagnetic interaction of the photons with
the electrons there arises the quantum transition between the photons and
the excitons. Applying the second quantization formalism to describe the
excitons in the local approximation, we denote $B_{\sigma }\left( \mathbf{p}%
\right) $ and $B_{\sigma }^{+}\left( \mathbf{p}\right) $ the destruction and
creation operators of the exciton with the momentum $\mathbf{p}$ and the
energy $E(\mathbf{p})$ in the spin state labeled also by the index $\sigma
=1,2$ defined such that the quantum mutual transition between the photon in
the polarization state with the index $\sigma _{1}$ and the exciton in the
spin state with the index $\sigma _{2}$ is allowed if $\sigma _{1}=\sigma
_{2}$ and forbidden if $\sigma _{1}\neq \sigma _{2}$. The Hamiltonian of the
photon-exciton system without the exciton-exciton interaction can be written
in the form

\begin{eqnarray}
H_{0} &=&\sum\limits_{\sigma }\sum\limits_{\mathbf{p}}\left\{ 
\begin{array}{r}
\end{array}
\omega \left( \mathbf{p}\right) \gamma _{\sigma }^{+}\left( \mathbf{p}%
\right) \gamma _{\sigma }\left( \mathbf{p}\right) +E\left( \mathbf{p}\right)
B_{\sigma }^{+}\left( \mathbf{p}\right) B_{\sigma }\left( \mathbf{p}\right)
\right.  \nonumber \\
&&\left. +\frac{g\left( \mathbf{p}\right) }{2}\left[ \gamma _{\sigma
}^{+}\left( \mathbf{p}\right) B_{\sigma }\left( \mathbf{p}\right) +B_{\sigma
}^{+}\left( \mathbf{p}\right) \gamma _{\sigma }\left( \mathbf{p}\right)
\right] \right\}  \label{119}
\end{eqnarray}
with some function $g\left( \mathbf{p}\right) $. The Coulomb interactions
between the electrons and the holes induce some effective exciton-exciton
direct- coupling with the interaction Hamiltonian

\begin{equation}
H_{\text{int}}=\sum\limits_{\sigma _{1}\sigma _{2}}\sum\limits_{\mathbf{p}%
_{1}\mathbf{p}_{2}}\sum\limits_{\mathbf{k}}B_{\sigma _{1}}^{+}\left( \mathbf{%
p}_{1}+\mathbf{k}\right) B_{\sigma _{2}}^{+}\left( \mathbf{p}_{2}-\mathbf{k}%
\right) \widetilde{U}_{\sigma _{1}\sigma _{2}}\left( \mathbf{k}\right)
B_{\sigma _{1}}\left( \mathbf{p}_{1}\right) B_{\sigma _{2}}\left( \mathbf{p}%
_{2}\right) .  \label{120}
\end{equation}
The Fourier transform $\widetilde{U}_{\sigma _{1}\sigma _{2}}\left( \mathbf{k%
}\right) $ of the effective interactions potential energies were given, for
example, in Refs$^{\left[ 22,23\right] }$. The functional integral of the
system of interacting photons and excitons is

\begin{eqnarray}
Z &=&\int \left[ D\gamma \right] \left[ D\gamma ^{+}\right] \left[ DB\right]
\left[ DB^{+}\right]  \nonumber \\
&&\exp \left\{ i\int dt\sum\limits_{\sigma }\sum\limits_{\mathbf{p}}\left(
\gamma _{\sigma }^{+}\left( \mathbf{p},t\right) \left[ i\frac{\partial }{%
\partial t}-\omega \left( \mathbf{p}\right) \right] \gamma _{\sigma }\left( 
\mathbf{p}\right) +B_{\sigma }^{+}\left( \mathbf{p,}t\right) \left[ i\frac{%
\partial }{\partial t}-E\left( \mathbf{p}\right) \right] B_{\sigma }\left( 
\mathbf{p,}t\right) \right. \right.  \nonumber \\
&&\left. \left. +\frac{g\left( \mathbf{p}\right) }{2}\left[ \gamma _{\sigma
}^{+}\left( \mathbf{p,}t\right) B_{\sigma }\left( \mathbf{p,}t\right)
+B_{\sigma }^{+}\left( \mathbf{p,}t\right) \gamma _{\sigma }\left( \mathbf{p,%
}t\right) \right] \right) \right\}  \label{121} \\
&&\exp \left\{ -\frac{i}{2}\sum\limits_{\sigma _{1}\sigma _{2}}\sum\limits_{%
\mathbf{p}_{1}\mathbf{p}_{2}}\sum\limits_{\mathbf{k}}B_{\sigma
_{1}}^{+}\left( \mathbf{p}_{1}+\mathbf{k,}t\right) B_{\sigma _{2}}^{+}\left( 
\mathbf{p}_{2}-\mathbf{k},t\right) \widetilde{U}_{\sigma _{1}\sigma
_{2}}\left( \mathbf{k}\right) B_{\sigma _{1}}\left( \mathbf{p}_{1},t\right)
B_{\sigma _{2}}\left( \mathbf{p}_{2},t\right) \right\} .  \nonumber
\end{eqnarray}
If we neglect the direct exciton-exciton coupling, then the functional
integral of the system becomes

\begin{eqnarray}
Z_{0} &=&\int \left[ D\gamma \right] \left[ D\gamma ^{+}\right] \left[
DB\right] \left[ DB^{+}\right]  \nonumber \\
&&\exp \left\{ i\int dt\sum\limits_{\sigma }\sum\limits_{\mathbf{p}}\left(
\gamma _{\sigma }^{+}\left( \mathbf{p},t\right) \left[ i\frac{\partial }{%
\partial t}-\omega \left( \mathbf{p}\right) \right] \gamma _{\sigma }\left( 
\mathbf{p}\right) +B_{\sigma }^{+}\left( \mathbf{p,}t\right) \left[ i\frac{%
\partial }{\partial t}-E\left( \mathbf{p}\right) \right] B_{\sigma }\left( 
\mathbf{p,}t\right) \right. \right.  \nonumber \\
&&\left. \left. +\frac{g\left( \mathbf{p}\right) }{2}\left[ \gamma _{\sigma
}^{+}\left( \mathbf{p,}t\right) B_{\sigma }\left( \mathbf{p,}t\right)
+B_{\sigma }^{+}\left( \mathbf{p,}t\right) \gamma _{\sigma }\left( \mathbf{p,%
}t\right) \right] \right) \right\}  \label{122}
\end{eqnarray}

In order to describe the bipolaritons we introduce the composite symmetric
bi-local field

\begin{equation}
\Phi _{\sigma _{1}\sigma _{2}}\left( \mathbf{p}_{1},\mathbf{p}_{2},t\right)
=\Phi _{\sigma _{2}\sigma _{1}}\left( \mathbf{p}_{2},\mathbf{p}_{1},t\right)
\label{123}
\end{equation}
depending on two momenta $\mathbf{p}_{1}$ and $\mathbf{p}_{2}$ and being
labeled by two indices $\sigma _{1},\sigma _{2}$. Denote $\Phi _{\sigma
_{1}\sigma _{2}}^{+}\left( \mathbf{p}_{1},\mathbf{p}_{2},t\right) $ its
hermitian conjugate and set

\begin{eqnarray}
Z_{0}^{\Phi \Phi ^{+}} &=&\int \left[ D\Phi \right] \left[ D\Phi ^{+}\right]
\label{124} \\
&&\exp \left\{ \frac{i}{2}\int dt\sum\limits_{\sigma _{1}\sigma
_{2}}\sum\limits_{\mathbf{p}_{1}\mathbf{p}_{2}}\sum\limits_{\mathbf{k}}\Phi
_{\sigma _{1}\sigma _{2}}^{+}\left( \mathbf{p}_{1}+\mathbf{k},\mathbf{p}_{2}-%
\mathbf{k},t\right) \widetilde{U}_{\sigma _{1}\sigma _{2}}\left( \mathbf{k}%
\right) \Phi _{\sigma _{1}\sigma _{2}}\left( \mathbf{p}_{1},\mathbf{p}%
_{2},t\right) \right\}  \nonumber
\end{eqnarray}
Shifting the functional integration variables

\begin{eqnarray}
\Phi _{\sigma _{1}\sigma _{2}}\left( \mathbf{p}_{1},\mathbf{p}_{2},t\right)
&\rightarrow &\Phi _{\sigma _{1}\sigma _{2}}\left( \mathbf{p}_{1},\mathbf{p}%
_{2},t\right) +B_{\sigma _{1}}\left( \mathbf{p}_{1}\mathbf{,}t\right)
B_{\sigma _{2}}\left( \mathbf{p}_{2}\mathbf{,}t\right) ,  \label{125} \\
\Phi _{\sigma _{1}\sigma _{2}}^{+}\left( \mathbf{p}_{1},\mathbf{p}%
_{2},t\right) &\rightarrow &\Phi _{\sigma _{1}\sigma _{2}}^{+}\left( \mathbf{%
p}_{1}+\mathbf{k},\mathbf{p}_{2}-\mathbf{k},t\right) +B_{\sigma
_{1}}^{+}\left( \mathbf{p}_{1}+\mathbf{k,}t\right) B_{\sigma _{2}}^{+}\left( 
\mathbf{p}_{2}-\mathbf{k,}t\right) ,  \nonumber
\end{eqnarray}

\noindent we establish the Hubbard-Stratonovich transformation

\begin{eqnarray}
&&\exp \left\{ -\frac{i}{2}\int dt\sum\limits_{\sigma _{1}\sigma
_{2}}\sum\limits_{\mathbf{p}_{1}\mathbf{p}_{2}}\sum\limits_{\mathbf{k}%
}B_{\sigma _{1}}^{+}\left( \mathbf{p}_{1}+\mathbf{k},t\right) B_{\sigma
_{2}}^{+}\left( \mathbf{p}_{2}-\mathbf{k},t\right) \widetilde{U}_{\sigma
_{1}\sigma _{2}}\left( \mathbf{k}\right) B_{\sigma _{1}}\left( \mathbf{p}%
_{1},t\right) B_{\sigma _{2}}\left( \mathbf{p}_{2},t\right) \right\} 
\nonumber \\
&=&\frac{1}{Z_{0}^{\Phi \Phi ^{+}}}\int \left[ D\Phi \right] \left[ D\Phi
^{+}\right]  \label{126} \\
&&.\exp \left\{ \frac{i}{2}\int dt\sum\limits_{\sigma _{1}\sigma
_{2}}\sum\limits_{\mathbf{p}_{1}\mathbf{p}_{2}}\sum\limits_{\mathbf{k}}\Phi
_{\sigma _{1}\sigma _{2}}^{+}\left( \mathbf{p}_{1}+\mathbf{k},\mathbf{p}_{2}-%
\mathbf{k},t\right) \widetilde{U}_{\sigma _{1}\sigma _{2}}\left( \mathbf{k}%
\right) \Phi _{\sigma _{1}\sigma _{2}}\left( \mathbf{p}_{1},\mathbf{p}%
_{2},t\right) \right\}  \nonumber \\
&&.\exp \left\{ \frac{i}{2}\int dt\sum\limits_{\sigma _{1}\sigma
_{2}}\sum\limits_{\mathbf{p}_{1}\mathbf{p}_{2}}\left[ \Delta _{\sigma
_{1}\sigma _{2}}^{+}\left( \mathbf{p}_{1},\mathbf{p}_{2},t\right) B_{\sigma
_{1}}\left( \mathbf{p}_{1},t\right) B_{\sigma _{2}}\left( \mathbf{p}%
_{2},t\right) \right. \right.  \nonumber \\
&&\,\,\,\,\,\,\,\,\,\,\,\,\,\,\,\,\,\,\left. \left. B_{\sigma
_{1}}^{+}\left( \mathbf{p}_{1},t\right) B_{\sigma _{2}}^{+}\left( \mathbf{p}%
_{2},t\right) \Delta _{\sigma _{1}\sigma _{2}}\left( \mathbf{p}_{1},\mathbf{p%
}_{2},t\right) \right] 
\begin{array}{r}
\end{array}
\right\} ,  \nonumber
\end{eqnarray}
where

\begin{eqnarray}
\Delta _{\sigma _{1}\sigma _{2}}\left( \mathbf{p}_{1},\mathbf{p}%
_{2},t\right) &=&\sum\limits_{\mathbf{k}}\widetilde{U}_{\sigma _{1}\sigma
_{2}}\left( \mathbf{k}\right) \Phi _{\sigma _{1}\sigma _{2}}\left( \mathbf{p}%
_{1}-\mathbf{k},\mathbf{p}_{2}+\mathbf{k},t\right) ,  \nonumber \\
\Delta _{\sigma _{1}\sigma _{2}}^{+}\left( \mathbf{p}_{1},\mathbf{p}%
_{2},t\right) &=&\sum\limits_{\mathbf{k}}\Phi _{\sigma _{1}\sigma
_{2}}^{+}\left( \mathbf{p}_{1}-\mathbf{k},\mathbf{p}_{2}+\mathbf{k},t\right) 
\widetilde{U}_{\sigma _{1}\sigma _{2}}\left( -\mathbf{k}\right) .
\label{127}
\end{eqnarray}
Using this relation to transform the last exponential in the r.h.s. of the
formula (121), reversing the order of the functional integrations over the
elementary fields $\gamma _{\sigma },\,\gamma _{\sigma }^{+},\,B_{\sigma
},\,B_{\sigma }^{+}$ and the composite ones $\Phi _{\sigma _{1}\sigma
_{2}},\,\Phi _{\sigma _{1}\sigma _{2}}^{+}$ and then integrating out over
the elementary fields, we rewrite $Z$ in the form of a functional integral
over the bi-local composite field $\Phi _{\sigma _{1}\sigma _{2}}$ and its
conjugate $\,\Phi _{\sigma _{1}\sigma _{2}}^{+}$

\begin{equation}
Z=\frac{Z_{0}}{Z_{0}^{\Phi \Phi ^{+}}}\int \left[ D\Phi \right] \left[ D\Phi
^{+}\right] \exp \left\{ iS_{\text{eff}}\left[ \Phi ,\Phi ^{+}\right]
\right\}  \label{128}
\end{equation}
with the effective action of the bi-local composite field

\begin{eqnarray}
S_{\text{eff}}\left[ \Phi ,\Phi ^{+}\right] &=&\frac{1}{2}\int
dt\sum\limits_{\sigma _{1}\sigma _{2}}\sum\limits_{\mathbf{p}_{1}\mathbf{p}%
_{2}}\sum\limits_{\mathbf{k}}  \label{129} \\
&&\Phi _{\sigma _{1}\sigma _{2}}^{+}\left( \mathbf{p}_{1}+\mathbf{k},\mathbf{%
p}_{2}-\mathbf{k},t\right) \widetilde{U}_{\sigma _{1}\sigma _{2}}\left( 
\mathbf{k}\right) \Phi _{\sigma _{1}\sigma _{2}}\left( \mathbf{p}_{1},%
\mathbf{p}_{2},t\right) +W\left[ \Delta ,\Delta ^{+}\right] ,  \nonumber
\end{eqnarray}
where the functional $W\left[ \Delta ,\Delta ^{+}\right] $ is determined by
the formula

\begin{eqnarray}
\exp \left\{ iW\left[ \Delta ,\Delta ^{+}\right] \right\} &=&\frac{1}{Z^{0}}%
\int \left[ D\gamma \right] \left[ D\gamma ^{+}\right] \left[ DB\right]
\left[ DB^{+}\right]  \nonumber \\
&&\exp \left\{ i\int dt\sum\limits_{\sigma }\sum\limits_{\mathbf{p}}\left(
\gamma _{\sigma }^{+}\left( \mathbf{p},t\right) \left[ i\frac{\partial }{%
\partial t}-\omega \left( \mathbf{p}\right) \right] \gamma _{\sigma }\left( 
\mathbf{p}\right) \right. \right.  \nonumber \\
&&\,\,\,\left. +B_{\sigma }^{+}\left( \mathbf{p,}t\right) \left[ i\frac{%
\partial }{\partial t}-E\left( \mathbf{p}\right) \right] B_{\sigma }\left( 
\mathbf{p,}t\right) \right.  \label{130} \\
&&\left. \left. +\frac{g\left( \mathbf{p}\right) }{2}\left[ \gamma _{\sigma
}^{+}\left( \mathbf{p,}t\right) B_{\sigma }\left( \mathbf{p,}t\right)
+B_{\sigma }^{+}\left( \mathbf{p,}t\right) \gamma _{\sigma }\left( \mathbf{p,%
}t\right) \right] \right) \right\}  \nonumber \\
&&.\exp \left\{ \frac{i}{2}\int dt\sum\limits_{\sigma _{1}\sigma
_{2}}\sum\limits_{\mathbf{p}_{1}\mathbf{p}_{2}}\left[ \Delta _{\sigma
_{1}\sigma _{2}}^{+}\left( \mathbf{p}_{1},\mathbf{p}_{2},t\right) B_{\sigma
_{1}}\left( \mathbf{p}_{1},t\right) B_{\sigma _{2}}\left( \mathbf{p}%
_{2},t\right) \right. \right.  \nonumber \\
&&\,\,\,\,\,\left. \left. 
\begin{array}{r}
\end{array}
+B_{\sigma _{1}}^{+}\left( \mathbf{p}_{1},t\right) B_{\sigma _{2}}^{+}\left( 
\mathbf{p}_{2},t\right) \Delta _{\sigma _{1}\sigma _{2}}\left( \mathbf{p}%
_{1},\mathbf{p}_{2},t\right) \right] \right\}  \nonumber
\end{eqnarray}
and can be represented in the form of a functional power series in the
fields $\Delta _{\sigma _{1}\sigma _{2}}\left( \mathbf{p}_{1},\mathbf{p}%
_{2},t\right) $ and $\Delta _{\sigma _{1}\sigma _{2}}^{+}\left( \mathbf{p}%
_{1},\mathbf{p}_{2},t\right) $

\begin{equation}
W\left[ \Delta ,\Delta ^{+}\right] =\sum\limits_{n=1}^{\infty }W^{\left(
2n\right) }\left[ \Delta ,\Delta ^{+}\right]  \label{131}
\end{equation}
$W^{\left( 2n\right) }\left[ \Delta ,\Delta ^{+}\right] $ being a
homogeneous functional of $n-th$ order with respect to each type of fields $%
\Delta _{\sigma _{1}\sigma _{2}}\left( \mathbf{p}_{1},\mathbf{p}%
_{2},t\right) $ and $\Delta _{\sigma _{1}\sigma _{2}}^{+}\left( \mathbf{p}%
_{1},\mathbf{p}_{2},t\right) .$

In order to calculate the functional integral in the formula (130) and
derive the explicit expressions of $W^{\left( 2n\right) }\left[ \Delta
,\Delta ^{+}\right] $ we diagonalize the Hamiltonian $H_{0}$ of the free
polaritons by means of the Bogolubov transformation

\begin{eqnarray}
B\left( \mathbf{p}\right) &=&u\left( \mathbf{p}\right) c_{1\sigma }+v\left( 
\mathbf{p}\right) c_{2\sigma }\left( \mathbf{p}\right) ,  \nonumber \\
\gamma _{\sigma }\left( \mathbf{p}\right) &=&-v\left( \mathbf{p}\right)
c_{1\sigma }+u\left( \mathbf{p}\right) c_{2\sigma }\left( \mathbf{p}\right) .
\label{132}
\end{eqnarray}
It can be shown that with the transformation coefficient $u\left( \mathbf{p}%
\right) $ and $v\left( \mathbf{p}\right) $ are determined by the relations

\begin{eqnarray}
u\left( \mathbf{p}\right) ^{2} &=&\frac{1}{2}\left\{ 1+\frac{\omega \left( 
\mathbf{p}\right) -E\left( \mathbf{p}\right) }{\left( \left[ \omega \left( 
\mathbf{p}\right) -E\left( \mathbf{p}\right) \right] ^{2}+g\left( \mathbf{p}%
\right) ^{2}\right) ^{1/2}}\right\} ,  \nonumber \\
v\left( \mathbf{p}\right) ^{2} &=&\frac{1}{2}\left\{ 1-\frac{\omega \left( 
\mathbf{p}\right) -E\left( \mathbf{p}\right) }{\left( \left[ \omega \left( 
\mathbf{p}\right) -E\left( \mathbf{p}\right) \right] ^{2}+g\left( \mathbf{p}%
\right) ^{2}\right) ^{1/2}}\right\} ,  \label{133} \\
2u\left( \mathbf{p}\right) v\left( \mathbf{p}\right) &=&\frac{g\left( 
\mathbf{p}\right) }{\left( \left[ \omega \left( \mathbf{p}\right) -E\left( 
\mathbf{p}\right) \right] ^{2}+g\left( \mathbf{p}\right) ^{2}\right) ^{1/2}},
\nonumber
\end{eqnarray}
the Hamiltonian $H_{0}$ becomes

\begin{equation}
H_{0}=\sum\limits_{i=1,2}\sum\limits_{\sigma }\sum\limits_{\mathbf{p}%
}E_{i}\left( p\right) c_{i\sigma }^{+}\left( \mathbf{p}\right) c_{i\sigma
}\left( \mathbf{p}\right) ,  \label{134}
\end{equation}
where

\begin{eqnarray}
E_{1}\left( \mathbf{p}\right) &=&\frac{1}{2}\left\{ \omega \left( \mathbf{p}%
\right) +E\left( \mathbf{p}\right) +\left( \left[ \omega \left( \mathbf{p}%
\right) -E\left( \mathbf{p}\right) \right] ^{2}+g\left( \mathbf{p}\right)
^{2}\right) ^{1/2}\right\} ,  \nonumber \\
E_{2}\left( \mathbf{p}\right) &=&\frac{1}{2}\left\{ \omega \left( \mathbf{p}%
\right) +E\left( \mathbf{p}\right) -\left( \left[ \omega \left( \mathbf{p}%
\right) -E\left( \mathbf{p}\right) \right] ^{2}+g\left( \mathbf{p}\right)
^{2}\right) ^{1/2}\right\} .  \label{135}
\end{eqnarray}
$c_{i\sigma }\left( \mathbf{p}\right) $ and $c_{i\sigma }^{+}\left( \mathbf{p%
}\right) $ are the destruction and creation operators of the polariton with
the momentum $\mathbf{p}$, energy $E_{i}\left( \mathbf{p}\right) $ and
polarization $\sigma $ in the branche $i$. In terms of the mew functional
integration variables we have

\begin{equation}
Z_{0}=\int \left[ Dc_{i}\right] \left[ Dc_{i}^{+}\right] \exp \left\{ i\int
dt\sum\limits_{i}\sum\limits_{\sigma }\sum\limits_{\mathbf{p}}c_{i\sigma
}^{+}\left( \mathbf{p},t\right) \left[ i\frac{\partial }{\partial t}%
-E_{i}\left( \mathbf{p}\right) \right] c_{i\sigma }\left( \mathbf{p}%
,t\right) \right\} .  \label{136}
\end{equation}
Define the two-point Green functions $G_{i}\left( \mathbf{p}%
,t_{1}-t_{2}\right) $ of the free polaritons in the following manner

\begin{eqnarray}
&&\frac{1}{Z_{0}}\int \left[ Dc_{i}\right] \left[ Dc_{i}^{+}\right]
c_{i_{1}\sigma _{1}}^{+}\left( \mathbf{p}_{1},t_{1}\right) c_{i_{2}\sigma
_{2}}^{+}\left( \mathbf{p}_{2},t_{2}\right)  \nonumber \\
&&\exp \left\{ i\int dt\sum\limits_{i}\sum\limits_{\sigma }\sum\limits_{%
\mathbf{p}}c_{i\sigma }^{+}\left( \mathbf{p},t\right) \left[ i\frac{\partial 
}{\partial t}-E_{i}\left( \mathbf{p}\right) \right] c_{i\sigma }\left( 
\mathbf{p},t\right) \right\}  \label{137} \\
&=&i\delta _{i_{1}i_{2}}\delta _{\sigma _{1}\sigma _{2}}\delta _{\mathbf{p}%
_{1}\mathbf{p}_{2}}G_{i}\left( \mathbf{p}_{i},t_{1}-t_{2}\right)  \nonumber
\end{eqnarray}
and denote $\widetilde{G}_{i}\left( \mathbf{p},\omega \right) $ their
Fourier transforms with respect to the time variable

\begin{equation}
G_{i}\left( \mathbf{p},t\right) =\frac{1}{2\pi }\int e^{-i\omega t}%
\widetilde{G}_{i}\left( \mathbf{p},\omega \right) d\omega  \label{138}
\end{equation}
Instead of the time-dependent bi-local fields $\Phi _{\sigma _{1}\sigma
_{2}}\left( \mathbf{p}_{1},\mathbf{p}_{2},t\right) $, $\Phi _{\sigma
_{1}\sigma _{2}}^{+}\left( \mathbf{p}_{1},\mathbf{p}_{2},t\right) $
and\linebreak $\Delta _{\sigma _{1}\sigma _{2}}\left( \mathbf{p}_{1},\mathbf{%
p}_{2},t\right) $ ,\thinspace $\Delta _{\sigma _{1}\sigma _{2}}^{+}\left( 
\mathbf{p}_{1},\mathbf{p}_{2},t\right) $ it is also convenient to work with
their Fourier transforms with respect to the time variable $\widetilde{\Phi }%
_{\sigma _{1}\sigma _{2}}\left( \mathbf{p}_{1},\mathbf{p}_{2},\omega \right) 
$, $\widetilde{\Phi }_{\sigma _{1}\sigma _{2}}^{+}\left( \mathbf{p}_{1},%
\mathbf{p}_{2},\omega \right) $ and\linebreak $\widetilde{\Delta }_{\sigma
_{1}\sigma _{2}}\left( \mathbf{p}_{1},\mathbf{p}_{2},\omega \right) $%
,\thinspace $\,\,\,\widetilde{\Delta }_{\sigma _{1}\sigma _{2}}^{+}\left( 
\mathbf{p}_{1},\mathbf{p}_{2},\omega \right) $

\begin{eqnarray}
\Phi _{\sigma _{1}\sigma _{2}}\left( \mathbf{p}_{1},\mathbf{p}_{2},t\right)
&=&\frac{1}{2\pi }\int e^{-i\omega t}\widetilde{\Phi }_{\sigma _{1}\sigma
_{2}}\left( \mathbf{p}_{1},\mathbf{p}_{2},\omega \right) d\omega ,  \nonumber
\\
\Delta _{\sigma _{1}\sigma _{2}}\left( \mathbf{p}_{1},\mathbf{p}%
_{2},t\right) &=&\frac{1}{2\pi }\int e^{-i\omega t}\widetilde{\Delta }%
_{\sigma _{1}\sigma _{2}}\left( \mathbf{p}_{1},\mathbf{p}_{2},\omega \right)
d\omega .  \label{139}
\end{eqnarray}
In the second order with respect to the fields $\widetilde{\Phi }_{\sigma
_{1}\sigma _{2}}\left( \mathbf{p}_{1},\mathbf{p}_{2},\omega \right) $, $%
\widetilde{\Phi }_{\sigma _{1}\sigma _{2}}^{+}\left( \mathbf{p}_{1},\mathbf{p%
}_{2},\omega \right) $ and $\widetilde{\Delta }_{\sigma _{1}\sigma
_{2}}\left( \mathbf{p}_{1},\mathbf{p}_{2},\omega \right) $ ,\thinspace $%
\,\,\,\widetilde{\Delta }_{\sigma _{1}\sigma _{2}}^{+}\left( \mathbf{p}_{1},%
\mathbf{p}_{2},\omega \right) $ the effective action (129) becomes

\begin{eqnarray}
S_{\text{eff}}\left[ \Phi ,\Phi ^{+}\right] &\approx &\frac{1}{4\pi }\int
d\omega \sum\limits_{\sigma _{1}\sigma _{2}}\sum\limits_{\mathbf{p}_{1}%
\mathbf{p}_{2}}\,\,\widetilde{\Delta }_{\sigma _{1}\sigma _{2}}^{+}\left( 
\mathbf{p}_{1},\mathbf{p}_{2},\omega \right)  \label{140} \\
&&.\left[ \widetilde{\Phi }_{\sigma _{1}\sigma _{2}}\left( \mathbf{p}_{1},%
\mathbf{p}_{2},\omega \right) -M\left( \mathbf{p}_{1},\mathbf{p}_{2},\omega
\right) \widetilde{\Delta }_{\sigma _{1}\sigma _{2}}\left( \mathbf{p}_{1},%
\mathbf{p}_{2},\omega \right) \right] ,  \nonumber
\end{eqnarray}
where

\begin{eqnarray}
M\left( \mathbf{p},\mathbf{q},\omega \right) &=&\frac{i}{2\pi }\int \left[
u\left( \mathbf{p}\right) ^{2}\widetilde{G}_{1}\left( \mathbf{p},\omega
-\omega ^{\prime }\right) +v\left( \mathbf{p}\right) ^{2}\widetilde{G}%
_{2}\left( \mathbf{p},\omega -\omega ^{\prime }\right) \right]  \nonumber \\
&&\,\,\,\,\,\,\,\,\,\,\,\,\,\,\,\,\,\,\,\left[ u\left( \mathbf{q}\right) ^{2}%
\widetilde{G}_{1}\left( \mathbf{q},\omega ^{\prime }\right) +v\left( \mathbf{%
q}\right) ^{2}\widetilde{G}_{2}\left( \mathbf{q},\omega ^{\prime }\right)
\right] d\omega ^{\prime }.  \label{141}
\end{eqnarray}
Using the formula 
\begin{equation}
\widetilde{G}_{i}\left( \mathbf{p},\omega \right) =\frac{1}{\omega
-E_{i}\left( \mathbf{p}\right) +i0}  \label{142}
\end{equation}
and performing the integration over the variables $\omega ,$ we obtain

\begin{eqnarray}
M\left( \mathbf{p},\mathbf{q},\omega \right) &=&\frac{u\left( \mathbf{p}%
\right) ^{2}u\left( \mathbf{q}\right) ^{2}}{\omega -E_{1}\left( \mathbf{p}%
\right) -E_{1}\left( \mathbf{q}\right) }+\frac{u\left( \mathbf{p}\right)
^{2}v\left( \mathbf{q}\right) ^{2}}{\omega -E_{1}\left( \mathbf{p}\right)
-E_{2}\left( \mathbf{q}\right) }  \nonumber \\
&&+\frac{v\left( \mathbf{p}\right) ^{2}u\left( \mathbf{q}\right) ^{2}}{%
\omega -E_{2}\left( \mathbf{p}\right) -E_{1}\left( \mathbf{q}\right) }+\frac{%
v\left( \mathbf{p}\right) ^{2}v\left( \mathbf{q}\right) ^{2}}{\omega
-E_{2}\left( \mathbf{p}\right) -E_{2}\left( \mathbf{q}\right) }.  \label{143}
\end{eqnarray}
From the expression (14) of the effective action and the variational
principle

\begin{equation}
\frac{\delta S_{\text{eff}}\left[ \Phi ,\Phi ^{+}\right] }{\delta \Delta
_{\sigma _{1}\sigma _{2}}^{+}\left( \mathbf{p}_{1},\mathbf{p}_{2},\omega
\right) }=0  \label{144}
\end{equation}
it follows the field equation

\begin{equation}
\widetilde{\Phi }_{\sigma _{1}\sigma _{2}}\left( \mathbf{p}_{1},\mathbf{p}%
_{2},\omega \right) -M\left( \mathbf{p}_{1},\mathbf{p}_{2},\omega \right) 
\widetilde{\Delta }_{\sigma _{1}\sigma _{2}}\left( \mathbf{p}_{1},\mathbf{p}%
_{2},\omega \right) =0  \label{145}
\end{equation}
or, in another form

\begin{equation}
L\left( \mathbf{p}_{1},\mathbf{p}_{2},\omega \right) \widetilde{\Phi }%
_{\sigma _{1}\sigma _{2}}\left( \mathbf{p}_{1},\mathbf{p}_{2},\omega \right)
-\sum\limits_{\mathbf{k}}\widetilde{U}_{\sigma _{1}\sigma _{2}}\left( 
\mathbf{k}\right) \widetilde{\Phi }_{\sigma _{1}\sigma _{2}}\left( \mathbf{p}%
_{1}-\mathbf{k},\mathbf{p}_{2}+\mathbf{k},\omega \right) =0  \label{146}
\end{equation}
with

\begin{equation}
L\left( \mathbf{p}_{1},\mathbf{p}_{2},\omega \right) =\frac{1}{M\left( 
\mathbf{p}_{1},\mathbf{p}_{2},\omega \right) }  \label{147}
\end{equation}
it becomes the Schrodinger equation for the biexcitons

\begin{eqnarray}
&&\left[ E_{1}\left( \mathbf{p}_{1}\right) +E_{2}\left( \mathbf{p}%
_{2}\right) \right] \widetilde{\Phi }_{\sigma _{1}\sigma _{2}}\left( \mathbf{%
p}_{1},\mathbf{p}_{2},\omega \right) +  \nonumber \\
&&+\sum\limits_{\mathbf{k}}\widetilde{U}_{\sigma _{1}\sigma _{2}}\left( 
\mathbf{k}\right) \widetilde{\Phi }_{\sigma _{1}\sigma _{2}}\left( \mathbf{p}%
_{1}-\mathbf{k},\mathbf{p}_{2}+\mathbf{k},\omega \right) 
\begin{array}{r}
=
\end{array}
\omega \widetilde{\Phi }_{\sigma _{1}\sigma _{2}}\left( \mathbf{p}_{1},%
\mathbf{p}_{2},\omega \right)  \label{148}
\end{eqnarray}
In terms of the functions of the space-time variables and the differential
operators of the space-time coordinates we can rewrite the Bethe-Salpeter
equation (146) in the form

\begin{equation}
\left[ L\left( -i\frac{\partial }{\partial \mathbf{r}_{1}},-i\frac{\partial 
}{\partial \mathbf{r}_{2}},i\frac{\partial }{\partial t}\right) -U_{\sigma
_{1}\sigma _{2}}\left( \mathbf{r}_{1}-\mathbf{r}_{2}\right) \right] \Phi
_{\sigma _{1}\sigma _{2}}\left( \mathbf{r}_{1},\mathbf{r}_{2},t\right) =0,
\label{149}
\end{equation}
where $U_{\sigma _{1}\sigma _{2}}\left( \mathbf{r}_{1}-\mathbf{r}_{2}\right) 
$ are the effective interaction potential energies between two excitons in
the polarization states $\sigma _{1}$ and $\sigma _{2}.$

\section{Conclusion and Discussions}

In this series of lectures we have presented the universal method based on
the application of functional integral technique for the study of the bound
states of the systems of two relativistic particles in high energy physics
or two quasiparticles with arbitrary dispersion laws in the condensed
matters. The Bethe-Salpeter equation for the composite fields describing the
corresponding two-body systems were derived. In the special case of the
direct local $4-$fermion coupling the Bethe-Salpeter equation reduces to the
Nambu-Jona-Lasinio equation. In the derivation of these equations we have
observed a very interesting phenomenon : there might exist some composite
boson fields with non-vanishing vacuum expectation values-the dynamical
Higgs fields. The existence of these fields and the spontaneous breaking of
the corresponding symmetries in QCD should be studied subsequently.

On the example of the bipolariton problem we have demonstrated how to apply
the functional integral method in order to derive the Bethe-Salpeter
equation for the bound states of two quasiparticles with the complicated
energy spectra in the condensed matters. The solution of the established
bound state equation for the bipolaritons, the derivation of the
Bethe-Salpeter equations for other two-body systems in the condensed
matters, for example the biphonons, the bimagnons, the electron-phonon or
electron-magnon bound states etc.. as well as the solution of these
equations would be also, the interesting subjects for the subsequent study.

The presented functional integral method can be easily extended for the
application to the study of different three-body systems in high energy
physics as well as in condensed matters theory. For the definiteness and as
an example let us consider the formation of the baryons as bound states of
the tree-quark systems - the triquark. As the necessary tool for the
application of the functional integral method with the use of the
Hubbard-Stratonovich transformation we introduce the six-quark (non-local,
in general) coupling induced by the interaction mechanism in QCD (instanton
induced, gluon exchange etc.) with the contribution

\begin{eqnarray}
S_{\text{int}}\left[ \psi ,\overline{\psi }\right] &=&\frac{1}{6}\int
d^{4}k_{1}\int d^{4}p_{1}\int d^{4}q_{1}\int d^{4}k_{2}\int d^{4}p_{2}\int
d^{4}q_{2}  \nonumber \\
&&\delta ^{\left( 4\right) }\left(
k_{1}+p_{1}+q_{1}-k_{2}-p_{2}-q_{2}\right) \widetilde{\overline{\psi }}%
^{A_{1}}\left( k_{1}\right) \widetilde{\overline{\psi }}^{B_{1}}\left(
p_{1}\right) \widetilde{\overline{\psi }}^{C_{1}}\left( q_{1}\right) \qquad
\qquad  \label{150} \\
&&V_{C_{1}B_{1}A_{1}}^{A_{2}B_{2}C_{2}}\left(
q_{1},p_{1},k_{1};k_{2},p_{2},q_{2}\right) \widetilde{\psi }_{C_{2}}\left(
q_{2}\right) \widetilde{\psi }_{B_{2}}\left( p_{2}\right) \widetilde{\psi }%
_{A_{2}}\left( k_{2}\right)  \nonumber
\end{eqnarray}
to the effective action reproducing the $6$-point Green functions of the
interacting quark field. In order to describe the baryons as the triquarks
we introduce the tri-local composite fermionic field and its conjugate with
the Fourier transforms $\widetilde{\Psi }_{CBA}\left( q,p,k\right) $ and $%
\widetilde{\overline{\Psi }}^{ABC}\left( k,p,q\right) $. We set

\begin{eqnarray}
Z_{0}^{\Psi ,\overline{\Psi }} &=&\int \left[ D\Psi \right] \left[ D%
\overline{\Psi }\right] \exp \left\{ -\frac{i}{6}\int d^{4}k_{1}\int
d^{4}p_{1}\int d^{4}q_{1}\right.  \nonumber \\
&&\left. \int d^{4}k_{2}\int d^{4}p_{2}\int d^{4}q_{2}\delta ^{\left(
4\right) }\left( k_{1}+p_{1}+q_{1}-k_{2}-p_{2}-q_{2}\right) \right.
\label{151} \\
&&\left. \widetilde{\overline{\Psi }}^{A_{1}B_{1}C_{1}}\left(
k_{1},p_{1},q_{1}\right) V_{C_{1}B_{1}A_{1}}^{A_{2}B_{2}C_{2}}\left(
q_{1},p_{1},k_{1};k_{2},p_{2},q_{2}\right) \widetilde{\Psi }%
_{C_{2}B_{2}A_{2}}\left( q_{2},p_{2},k_{2}\right) \right\}  \nonumber
\end{eqnarray}

and establish the Hubbard-Stratonovich transformation

\begin{eqnarray}
&&\exp \left\{ \frac{i}{6}\int d^{4}k_{1}\int d^{4}p_{1}\int d^{4}q_{1}\int
d^{4}k_{2}\int d^{4}p_{2}\int d^{4}q_{2}\delta ^{\left( 4\right) }\left(
k_{1}+p_{1}+q_{1}-k_{2}-p_{2}-q_{2}\right) \right.  \nonumber \\
&&\left. \widetilde{\overline{\psi }}^{A_{1}}\left( k_{1}\right) \widetilde{%
\overline{\psi }}^{B_{1}}\left( p_{1}\right) \widetilde{\overline{\psi }}%
^{C_{1}}\left( q_{1}\right) V_{C_{1}B_{1}A_{1}}^{A_{2}B_{2}C_{2}}\left(
q_{1},p_{1},k_{1};k_{2},p_{2},q_{2}\right) \widetilde{\psi }_{C_{2}}\left(
q_{2}\right) \widetilde{\psi }_{B_{2}}\left( p_{2}\right) \widetilde{\psi }%
_{A_{2}}\left( k_{2}\right) \right\}  \nonumber \\
&=&\frac{1}{Z_{0}^{\Psi ,\overline{\Psi }}}\int \left[ D\Psi \right] \left[ D%
\overline{\Psi }\right] \exp \left\{ -\frac{i}{6}\int d^{4}k_{1}\int
d^{4}p_{1}\int d^{4}q_{1}\right.  \label{152} \\
&&\,\,\left. \,\,\,\int d^{4}k_{2}\int d^{4}p_{2}\int d^{4}q_{2}\delta
^{\left( 4\right) }\left( k_{1}+p_{1}+q_{1}-k_{2}-p_{2}-q_{2}\right) \right.
\nonumber \\
&&\,\left. \widetilde{\overline{\Psi }}^{A_{1}B_{1}C_{1}}\left(
k_{1},p_{1},q_{1}\right) V_{C_{1}B_{1}A_{1}}^{A_{2}B_{2}C_{2}}\left(
q_{1},p_{1},k_{1};k_{2},p_{2},q_{2}\right) \widetilde{\Psi }%
_{C_{2}B_{2}A_{2}}\left( q_{2},p_{2},k_{2}\right) \right\}  \nonumber \\
&&\exp \left\{ -\frac{i}{6}\int d^{4}k\int d^{4}p\int d^{4}q\right. 
\nonumber \\
&&\,\,\left. \left[ \widetilde{\overline{\psi }}^{A}\left( k\right) 
\widetilde{\overline{\psi }}^{B}\left( p\right) \widetilde{\overline{\psi }}%
^{C}\left( q\right) \widetilde{\Delta }_{CBA}\left( q,p,k\right) +\widetilde{%
\overline{\Delta }}^{ABC}\left( k,p,q\right) \widetilde{\psi }_{C}\left(
q\right) \widetilde{\psi }_{B}\left( p\right) \widetilde{\psi }_{A}\left(
k\right) \right] \right\} ,  \nonumber
\end{eqnarray}
where

\begin{eqnarray}
\widetilde{\Delta }_{CBA}\left( q,p,k\right) &=&\int d^{4}k^{\prime }\int
d^{4}p^{\prime }\int d^{4}q^{\prime }\delta ^{\left( 4\right) }\left(
k+p+q-k^{\prime }-p^{\prime }-q^{\prime }\right)  \nonumber \\
&&\,\,\,V_{CBA}^{A^{\prime }B^{\prime }C^{\prime }}\left( q,p,k;k^{\prime
},p^{\prime },q^{\prime }\right) \widetilde{\Psi }_{C^{\prime }B^{\prime
}A^{\prime }}\left( q^{\prime },p^{\prime },k^{\prime }\right) ,  \label{153}
\\
\widetilde{\overline{\Delta }}^{ABC}\left( k,p,q\right) &=&\int
d^{4}k^{\prime }\int d^{4}p^{\prime }\int d^{4}q^{\prime }\delta ^{\left(
4\right) }\left( k+p+q-k^{\prime }-p^{\prime }-q^{\prime }\right)  \nonumber
\\
&&\widetilde{\overline{\Psi }}^{A^{\prime }B^{\prime }C^{\prime }}\left(
k^{\prime },p^{\prime },q^{\prime }\right) V_{C^{\prime }B^{\prime
}A^{\prime }}^{ABC}\left( q^{\prime },p^{\prime },k^{\prime };k,p,q\right) .
\nonumber
\end{eqnarray}
Substituting the expression (152) into the r.h.s of the formula (4),
reversing the order of the functional integrations and integrating out over
the quark field and its conjugate, we rewrite the functional (4) in the new
form

\begin{equation}
Z=\frac{Z_{0}}{Z_{0}^{\Psi ,\overline{\Psi }}}\int \left[ D\Psi \right]
\left[ D\overline{\Psi }\right] \exp \left\{ S_{\text{eff}}\left[ \Psi ,%
\overline{\Psi }\right] \right\}  \label{154}
\end{equation}
with the effective action 
\begin{eqnarray}
S_{\text{eff}}\left[ \Psi ,\overline{\Psi }\right] &=&-\frac{1}{6}\int
d^{4}k_{1}\int d^{4}p_{1}\int d^{4}q_{1}\int d^{4}k_{2}\int d^{4}p_{2}\int
d^{4}q_{2}  \nonumber \\
&&.\delta ^{\left( 4\right) }\left(
k_{1}+p_{1}+q_{1}-k_{2}-p_{2}-q_{2}\right) \widetilde{\overline{\Psi }}%
^{A_{1}B_{1}C_{1}}\left( k_{1},p_{1},q_{1}\right) \\
&&.V_{C_{1}B_{1}A_{1}}^{A_{2}B_{2}C_{2}}\left(
q_{1},p_{1},k_{1};k_{2},p_{2},q_{2}\right) \widetilde{\Psi }%
_{C_{2}B_{2}A_{2}}\left( q_{2},p_{2},k_{2}\right) +W\left[ \Delta ,\overline{%
\Delta }\right] ,  \nonumber
\end{eqnarray}
$W\left[ \Delta ,\overline{\Delta }\right] $ being a functional power series
in the composite fields $\widetilde{\Delta }_{CBA}\left( q,p,k\right) $
and\linebreak $\widetilde{\overline{\Delta }}^{ABC}\left( k,p,q\right) .$ In
the lowest (second) order approximation we have

\begin{eqnarray}
W\left[ \Delta ,\overline{\Delta }\right] &\approx &W^{\left( 2\right)
}\left[ \Delta ,\overline{\Delta }\right] =\frac{1}{6}\int d^{4}k\int
d^{4}p\int d^{4}q  \label{156} \\
&&\widetilde{\overline{\Delta }}^{A_{1}B_{1}C_{1}}\left( k,p,q\right)
S_{C_{1}}^{C_{2}}\left( q\right) S_{B_{1}}^{B_{2}}\left( p\right)
S_{A_{1}}^{A_{2}}\left( k\right) \widetilde{\Delta }_{C_{2}B_{2}A_{2}}\left(
q,p,k\right) .  \nonumber
\end{eqnarray}
In this approximation from the variational principle

\begin{equation}
\frac{\delta S_{\text{eff}}\left[ \Psi ,\overline{\Psi }\right] }{\delta 
\widetilde{\overline{\Delta }}^{ABC}\left( k,p,q\right) }=0  \label{157}
\end{equation}
it follows the Bethe-Salpeter equation for the triquarks

\begin{eqnarray}
\widetilde{\Delta }_{CBA}\left( q,p,k\right) &=&\int d^{4}k^{\prime }\int
d^{4}p^{\prime }\int d^{4}q^{\prime }\delta \left( k+p+q-k^{\prime
}-p^{\prime }-q^{\prime }\right)  \label{158} \\
&&V_{CBA}^{A^{\prime }B^{\prime }C^{\prime }}\left( q,p,k;k^{\prime
},p^{\prime },q^{\prime }\right) S_{C_{1}}^{C_{2}}\left( q^{\prime }\right)
S_{B_{1}}^{B_{2}}\left( p^{\prime }\right) S_{A_{1}}^{A_{2}}\left( k^{\prime
}\right) \widetilde{\Delta }_{C_{2}B_{2}A_{2}}\left( q^{\prime },p^{\prime
},k^{\prime }\right) .  \nonumber
\end{eqnarray}
In the special case of the local direct $6-$fermion coupling of the quark
field the potentials $V_{CBA}^{A^{\prime }B^{\prime }C^{\prime }}\left(
q,p,k;k^{\prime },p^{\prime },q^{\prime }\right) $ do not depend on the
momenta $k,\,p,\,q,\,k^{\prime },\,p^{\prime },\,q^{\prime }\,\,\,$and the
Bethe-Salpeter equation (158) reduces to the Nambu-Jona-Lasinion established
and discussed in Ref$^{\left[ 32\right] }.$ They would be the basic
equations for the theoretical study of the baryon structure in QCD.

\vskip 0.4cm%
{\Large \bf Acknowledgements.}%

\vskip 0.3cm%
The author would like to express his sincere appreciation to the National
Natural Sciences Council of Vietnam for the support to this work.

\vskip 1.5cm
\begin{center}
{\Large \bf References.}%

\vskip 0.5cm%
\end{center}

\begin{itemize}
\item[{\lbrack 1]}]  Salpeter and H. Bethe, \textit{Phys. Rev}. \textbf{84}
(1951) 1232.

\item[{\lbrack 2]}]  P. Jain and H. J. Munczek,\textit{\ Phys. Rev.} \textbf{%
D44} (1991)1873; \textbf{D46} (1992)438; \textbf{D48} (1993)5403.

\item[{\lbrack 3]}]  M. Beyer, Phys. Rev. D58 (1998)053001.

\item[{\lbrack 4]}]  A. Abd El-Hady, A. Datta and J. P. Vary, \textit{Phys.
Rev}. \textbf{D58} (1998)014007.

\item[{\lbrack 5]}]  V. A. Miransky, I. A. Showkovy and L. C. R.
Wijewardhana, \textit{hep-ph/0003327, hep-ph/0009173}.

\item[{\lbrack 6]}]  I. A. Showkovy, \textit{nul-th/0010021}.

\item[{\lbrack 7]}]  Y. Nambu and G. Jona-Lasinio, \textit{Phys. Rev}. 
\textbf{123} (1960)345; \textbf{124} (1961)246.

\item[{\lbrack 8]}]  T. Eguchi and H. Sugawara, \textit{Phys. Rev}. \textbf{%
D10} (1974)4257.

\item[{\lbrack 9]}]  T. Eguchi, \textit{Phys. Rev}. \textbf{D14 }(1976)2755.

\item[{\lbrack 10]}]  K. Kikkawa, \textit{Prog. Theor. Phys.} \textbf{56}
(1976)947.

\item[{\lbrack 11]}]  M. K. Volkov, \textit{Ann. Phys.} \textbf{157}
(1984)282.

\item[{\lbrack 12]}]  S. Klimt, M. Lutz, U. Vogl and W. Weise, \textit{Nucl.
Phys}. \textbf{A516} (1990)429.

\item[{\lbrack 13]}]  V. Bernard and U. Meissner, \textit{Nucl. Phys}. 
\textbf{A489} (1998)647.

\item[{\lbrack 14]}]  S. P. Klevansky, \textit{Rev. Mod. Phys.} \textbf{64}
(1992)649.

\item[{\lbrack 15]}]  D. Ebert, H. Reinhardt and M. K. Volkov, \textit{Prog.
Part. Nucl. Phys.} \textbf{33} (1994)1.

\item[{\lbrack 16]}]  M. Alford, K. Rajagopal and F. Wilczek , \textit{Phys.
Lett. }\textbf{B422} (1998) 247; \textit{Nucl. Phys},. \textbf{B537} (1999)
443.

\item[{\lbrack 17]}]  T. Sch\"{a}fer and F. Wilczek, \textit{Phys. Rev. Lett.%
} \textbf{82} (1999) 3956; \textit{Phys. Lett.} \textbf{B450} (1999) 325.

\item[{\lbrack 18]}]  R. Rapp, T. Sch\"{a}fer, E. Shuryak and M. Velkovsky, 
\textit{Phys. Rev. Lett.},\textbf{\ 81} (1998) 53.

\item[{\lbrack 19]}]  N.Evans, S. Hsu and M. Schwetz, \textit{Nucl. Phys.} 
\textbf{B551} (1999) 275; \textit{Phys. Lett.} \textbf{B449} (1999) 281.

\item[{\lbrack 20].}]  D. T. Son, \textit{Phys. Rev.} \textbf{D59 }(1998)
094019.

\item[{\lbrack 21]}]  G. W. Carter and D. Diakonov, \textit{Phys. Rev}. 
\textbf{D60 }(1999) 01004.

\item[{\lbrack 22]}]  A. L. Ivanov, M. Hasuo, N. Nagasawa and H. Haug, 
\textit{Phy. Rev}. \textbf{B52} (1995)11017.

\item[{\lbrack 23]}]  Z. G. Koinov, \textit{Jour. Phys: Condens Matter,} 
\textbf{10} (1998)2389.

\item[{\lbrack 24]}]  P. Ramond, \textit{Field Theory: A Modern Primer},
Benjamin, New York 1981.

\item[{\lbrack 25]}]  B. Sakita, \textit{Quantum Theory of Many-Variable
Systems and Fields}, World Scientific, Singapore 1985.

\item[{\lbrack 26]}]  S. Weinberg, \textit{Nucl. Phys.} \textbf{B413}
(1994)567.

\item[{\lbrack 27]}]  Ch. G. van Weert,\textit{\ Proc. 2nd Workshop on
Thermal Field Theories and Their Applications,} Tsukuba, Japan, 23-17 July
1990, Eds. H. Ezawa, T. Arimitsu and Y. Hashimoto, North-Holland, Amsterdam
1991, p. 17-26.

\item[{\lbrack 28]}]  Nguyen Van Hieu., Functional Integral Techniques in
Condensed Matter Physics, \textit{in the book}: \textit{Computational
Approaches for Novel Condensed Matter Systems}, Ed. Mukunda Das, Plenum
Press, New York, 1994, p. 194-234.

\item[{\lbrack 29]}]  Nguyen Van Hieu, Lectures given at 5$^{th}$
International School in Theoretical Physics, Hanoi 27 December 1998-10
January 1999, \textit{Adv. Nat. Sci}., \textbf{2 }(2001)3; Lectures given at
the 6$^{th}$ International School in Theoretical Physics, Vung Tau 27
December 1999-10 January 2000, \textit{Adv. Nat. Sci}. \textbf{2} (2001).

\item[{\lbrack 30]}]  M. Moachiya, M. Ukita and R. Fukuda, \textit{Phys. Rev}%
. \textbf{D40} (1989)2654.

\item[{\lbrack 31]}]  R. Rapp, T. Sch\"{a}fer, E. Shuryak and M. Velkovsky,
preprint IASSNS-HEP-99/40, Princeton, 1999, \textit{hep-ph/9904353.}

\item[{\lbrack 32]}]  Nguyen Van Hieu, Proc. International Conference ''%
\textit{Physics at the Frontiers of the Standard Model}'', Edition
Fronti\`{e}res, Paris 1996, p. 135-142.
\end{itemize}

\end{document}